\documentclass[twocolumn]{aastex63}
\usepackage{graphicx}
\usepackage{amsmath}
\usepackage{amssymb}
\usepackage{cases}
\usepackage{mathrsfs}
\usepackage[flushleft]{threeparttable}
\usepackage{amssymb}
\usepackage{pifont}
\usepackage{amsmath}
\usepackage{epstopdf}
\usepackage{xcolor}
\usepackage[all]{hypcap}
\usepackage{mwe,tikz}
\usepackage{bm}
\usepackage{tabularx}
\usepackage{multirow}
\usepackage{booktabs}
\usepackage{epstopdf}
\usepackage{subfigure}
\usepackage{xspace}
\usepackage{rotating}
\usepackage{CJKutf8}

\newcommand{\fermi}{{\em Fermi}\xspace}

\shorttitle{GRB 210121A Detected by Two Small Missions}
\shortauthors{Wang et al.}

\begin{document}

\begin{CJK*}{UTF8}{gbsn}

\title{GRB 210121A: A Typical Fireball Burst Detected by Two Small Missions}

\correspondingauthor{Bin-Bin Zhang, Ming Zeng, Shao-Lin Xiong}
\email{bbzhang@nju.edu.cn,zengming@tsinghua.edu.cn,xiongsl@ihep.ac.cn}

 \author[0000-0002-9738-1238]{Xiangyu Ivy Wang (王翔煜)}
\affiliation{School of Astronomy and Space Science, Nanjing University, Nanjing 210093, China}
 
\author{Xutao Zheng}
\affiliation{Department of Engineering Physics, Tsinghua University, Beijing 100084, China}

\author{Shuo Xiao}
\affiliation{Key Laboratory of Particle Astrophysics, Institute of High Energy Physics, Chinese Academy of Sciences, 19B Yuquan Road, Beijing 100049, People’s Republic of China} 
\affil{University of Chinese Academy of Sciences, Chinese Academy of Sciences, Beijing 100049, China}

\author[0000-0002-5485-5042]{Jun Yang}
\affiliation{School of Astronomy and Space Science, Nanjing
University, Nanjing 210093, China}

\author[0000-0002-5550-4017]{Zi-Ke Liu}
\affiliation{School of Astronomy and Space Science, Nanjing
University, Nanjing 210093, China}

\author[0000-0003-0691-6688]{Yu-Han Yang}
\affiliation{School of Astronomy and Space Science, Nanjing
University, Nanjing 210093, China}

\author{Jin-Hang Zou}
\affiliation{College of Physics, Hebei Normal University, Shijiazhuang 050024, China}

\author[0000-0003-4111-5958]{Bin-Bin Zhang}
\affiliation{School of Astronomy and Space Science, Nanjing
University, Nanjing 210093, China}
\affiliation{Key Laboratory of Modern Astronomy and Astrophysics (Nanjing University), Ministry of Education, China}
\affiliation{Department of Physics and Astronomy, University of Nevada Las Vegas, NV 89154, USA}

\author{Ming Zeng}
\affiliation{Key Laboratory of Particle and Radiation Imaging (Tsinghua University), Ministry of Education, Beijing 100084, China}
\affiliation{Department of Engineering Physics, Tsinghua University, Beijing 100084, China}

\author{Shao-Lin Xiong}
\affiliation{Key Laboratory of Particle Astrophysics, Institute of High Energy Physics, Chinese Academy of Sciences, 19B Yuquan Road, Beijing 100049, People’s Republic of China}

\author{Hua Feng}
\affiliation{Department of Astronomy, Tsinghua University, Beijing 100084, China}

\author{Xin-Ying Song}
\affiliation{Key Laboratory of Particle Astrophysics, Institute of High Energy Physics, Chinese Academy of Sciences, 19B Yuquan Road, Beijing 100049, People’s Republic of China}

\author{Jiaxing Wen}
\affiliation{Department of Engineering Physics, Tsinghua University, Beijing 100084, China}

\author{Dacheng Xu}
\affiliation{Department of Engineering Physics, Tsinghua University, Beijing 100084, China}

\author{Guo-Yin Chen}
\affiliation{School of Astronomy and Space Science, Nanjing
University, Nanjing 210093, China}
\affiliation{Key Laboratory of Modern Astronomy and Astrophysics (Nanjing University), Ministry of Education, China}

\author{Yang Ni}
\affiliation{School of Astronomy and Space Science, Nanjing
University, Nanjing 210093, China}
\affiliation{Key Laboratory of Modern Astronomy and Astrophysics (Nanjing University), Ministry of Education, China}

\author{Zi-Jian Zhang}
\affiliation{School of Astronomy and Space Science, Nanjing
University, Nanjing 210093, China}
\affiliation{Key Laboratory of Modern Astronomy and Astrophysics (Nanjing University), Ministry of Education, China} 

\author{Yu-Xuan Wu}
\affiliation{School of Astronomy and Space Science, Nanjing
University, Nanjing 210093, China}
\affiliation{Key Laboratory of Modern Astronomy and Astrophysics (Nanjing University), Ministry of Education, China}

\author{Ce Cai}
\affiliation{Key Laboratory of Particle Astrophysics, Institute of High Energy Physics, Chinese Academy of Sciences, 19B Yuquan Road, Beijing 100049, People’s Republic of China}
\affil{University of Chinese Academy of Sciences, Chinese Academy of Sciences, Beijing 100049, China}

\author{Jirong Cang}
\affiliation{Department of Astronomy, Tsinghua University, Beijing 100084, China}

\author{Yun-Wei Deng}
\affiliation{School of Astronomy and Space Science, Nanjing
University, Nanjing 210093, China}
\affiliation{Key Laboratory of Modern Astronomy and Astrophysics (Nanjing University), Ministry of Education, China} 

\author{Huaizhong Gao}
\affiliation{Department of Engineering Physics, Tsinghua University, Beijing 100084, China}

\author{De-Feng Kong}
\affiliation{Guangxi Key Laboratory for Relativistic Astrophysics, School of Physical Science and Technology, Guangxi University, Nanning 530004 China}

\author{Yue Huang}
\affiliation{Key Laboratory of Particle Astrophysics, Institute of High Energy Physics, Chinese Academy of Sciences, 19B Yuquan Road, Beijing 100049, People’s Republic of China}

\author{Cheng-kui Li}
\affiliation{Key Laboratory of Particle Astrophysics, Institute of High Energy Physics, Chinese Academy of Sciences, 19B Yuquan Road, Beijing 100049, People’s Republic of China}
 
\author{Hong Li}
\affiliation{Department of Astronomy, Tsinghua University, Beijing 100084, China}

\author{Xiao-Bo Li}
\affiliation{Key Laboratory of Particle Astrophysics, Institute of High Energy Physics, Chinese Academy of Sciences, 19B Yuquan Road, Beijing 100049, People’s Republic of China}

\author{En-Wei Liang}
\affiliation{Guangxi Key Laboratory for Relativistic Astrophysics, School of Physical Science and Technology, Guangxi University, Nanning 530004 China}

\author[0000-0002-0633-5325]{Lin Lin}
\affiliation{Department of Astronomy, Beijing Normal University, Beijing 100875, China}

\author{Yihui Liu}
\affiliation{Department of Engineering Physics, Tsinghua University, Beijing 100084, China}

\author{Xiangyun Long}
\affiliation{Department of Astronomy, Tsinghua University, Beijing 100084, China}

\author{Dian Lu}
\affiliation{Key Laboratory of Particle and Radiation Imaging (Tsinghua University), Ministry of Education, Beijing 100084, China}
\affiliation{Department of Engineering Physics, Tsinghua University, Beijing 100084, China}

\author{Qi Luo}
\affiliation{Key Laboratory of Particle Astrophysics, Institute of High Energy Physics, Chinese Academy of Sciences, 19B Yuquan Road, Beijing 100049, People’s Republic of China}
\affil{University of Chinese Academy of Sciences, Chinese Academy of Sciences, Beijing 100049, China}

\author{Yong-Chang Ma}
\affiliation{School of Astronomy and Space Science, Nanjing
University, Nanjing 210093, China}
\affiliation{Key Laboratory of Modern Astronomy and Astrophysics (Nanjing University), Ministry of Education, China} 

\author[0000-0002-5550-4017]{Yan-Zhi Meng}
\affiliation{School of Astronomy and Space Science, Nanjing
University, Nanjing 210093, China}
\affiliation{Key Laboratory of Modern Astronomy and Astrophysics (Nanjing University), Ministry of Education, China}

\author{Wen-Xi Peng}
\affiliation{Key Laboratory of Particle Astrophysics, Institute of High Energy Physics, Chinese Academy of Sciences, 19B Yuquan Road, Beijing 100049, People’s Republic of China}

\author{Rui Qiao}
\affiliation{Key Laboratory of Particle Astrophysics, Institute of High Energy Physics, Chinese Academy of Sciences, 19B Yuquan Road, Beijing 100049, People’s Republic of China}

\author{Li-Ming Song}
\affiliation{Key Laboratory of Particle Astrophysics, Institute of High Energy Physics, Chinese Academy of Sciences, 19B Yuquan Road, Beijing 100049, People’s Republic of China}

\author{Yang Tian}
\affiliation{Key Laboratory of Particle and Radiation Imaging (Tsinghua University), Ministry of Education, Beijing 100084, China}
\affiliation{Department of Engineering Physics, Tsinghua University, Beijing 100084, China}
 
\author{Pei-Yuan Wang}
\affiliation{Kuang Yaming Honor School, Nanjing
University, Nanjing 210093, China}
 
\author{Ping Wang}
\affiliation{Key Laboratory of Particle Astrophysics, Institute of High Energy Physics, Chinese Academy of Sciences, 19B Yuquan Road, Beijing 100049, People’s Republic of China} 
 
\author{Xiang-Gao Wang}
\affiliation{Guangxi Key Laboratory for Relativistic Astrophysics, School of Physical Science and Technology, Guangxi University, Nanning 530004 China}

\author{Sheng Xu}
\affiliation{School of Astronomy and Space Science, Nanjing
University, Nanjing 210093, China}
\affiliation{Key Laboratory of Modern Astronomy and Astrophysics (Nanjing University), Ministry of Education, China} 

\author{Dongxin Yang}
\affiliation{Department of Engineering Physics, Tsinghua University, Beijing 100084, China}

\author{Yi-Han Yin}
\affiliation{School of Physics, Nanjing
University, Nanjing 210093, China}

\author{Weihe Zeng}
\affiliation{Department of Engineering Physics, Tsinghua University, Beijing 100084, China}

\author{Zhi Zeng}
\affiliation{Key Laboratory of Particle and Radiation Imaging (Tsinghua University), Ministry of Education, Beijing 100084, China}
\affiliation{Department of Engineering Physics, Tsinghua University, Beijing 100084, China}

\author{Ting-Jun Zhang}
\affiliation{Kuang Yaming Honor School, Nanjing
University, Nanjing 210093, China}
\affiliation{Key Laboratory of Modern Astronomy and Astrophysics (Nanjing University), Ministry of Education, China} 

\author{Yuchong Zhang}
\affiliation{Department of Physics, Tsinghua University, Beijing 100084, China}

\author[0000-0002-5485-5042]{Zhao Zhang}
\affiliation{School of Astronomy and Space Science, Nanjing
University, Nanjing 210093, China}
\affiliation{Key Laboratory of Modern Astronomy and Astrophysics (Nanjing University), Ministry of Education, China}

\author{Zhen Zhang}
\affiliation{Key Laboratory of Particle Astrophysics, Institute of High Energy Physics, Chinese Academy of Sciences, 19B Yuquan Road, Beijing 100049, People’s Republic of China}

\begin{abstract}

The Chinese CubeSat Mission, Gamma Ray Integrated Detectors (GRID), recently detected its first gamma-ray burst, GRB 210121A, {which was jointly observed by the Gravitational wave high-energy Electromagnetic Counterpart All-sky Monitor (GECAM)}. This burst is confirmed by several other missions, including \fermi and \textit{Insight}-HXMT. We combined multi-mission observational data and performed a comprehensive analysis of the burst's temporal and spectral properties. Our results show that the burst is relatively special in its high peak energy, thermal-like low energy indices, and large fluence. By putting it to the $E_{\rm p}$-$E_{\rm\gamma, iso}$ relation diagram with assumed distance, we found this burst can be constrained at the redshift range of [0.3,3.0]. The thermal spectral component is also confirmed by the direct fit of the physical models to the observed spectra. Interestingly, the physical photosphere model also constrained a redshift of $z\sim$ 0.3 for this burst, which help us to identify a host galaxy candidate at such a distance within the location error box. Assuming the host galaxy is real, we found the burst can be best explained by the photosphere emission of a typical fireball with an initial radius of $r_0\sim$ 3.2 $\times 10^7$ cm.  
\end{abstract}


\section{Introduction} \label{sec:intro}
\end{CJK*}

The radiation mechanism that powers the prompt emission of gamma-ray bursts (GRBs) remains controversial. The leading two models, namely the photospheric emission of the relativistic fireball not far from the central engine \citep{2008ApJ...682..463P,2011ApJ...726...90Z,2018ApJS..234....3G,2018ApJ...866...13H,2018ApJ...860...72M,2019ApJ...882...26M,2018NatAs...2...69Z} and the synchrotron emission of the Poynting flux of a magnetic outflow generated at a large radius \citep{Zhang_2009, 2011ApJ...726...90Z, 2016ApJ...816...72Z, 2018NatAs...2...69Z}, are both successful in reproducing some observed spectra on a burst-by-burst basis. The superposition of the thermal and the non-thermal components in some GRBs \citep{2011ApJ...727L..33G, Gao_2015} further complicates the diversity and points towards a ``hybrid" model. In this paper, we report a recent GRB 210121A, which contributes an additional smoking case that puts strong evidence onto the photospheric emission origin.

GRB 210121A was detected by two small GRB missions recently launched in China. In particular, GRID is a low-cost student project aiming to build an all-sky and full-time CubeSat network in low Earth orbits in the energy range from 20 keV to 2 MeV \citep{2019ExA....48...77W}, with a dedicated scientific goal of observing and accumulating a considerable large sample of GRBs. The first and the second CubeSats of GRID were successfully launched in 2018 and 2020, respectively. To date, GRID have detected dozens of GRB candidates and one confirmed burst, GRB 210121A. 
{Launched in Dec. 2020, GECAM \citep[][]{GCN29347} is a new Chinese high energy astrophysics mission consisting of two micro-satellites that aims to monitor all kinds of X-ray and gamma-ray transients in the energy range from about 6 keV to 5000 keV \citep{2020SSPMA..50l9512C}.}

In this paper, we performed a comprehensive analysis of the high-energy data of GRB 210121A observed by multiple missions (\S \ref{sec:data_ana}). Motivated by its high peak spectral energy and unusually hard spectral indices, we further investigated how particular the burst is by comparing its temporal and spectral properties to those of a large GRB sample and by placing the burst onto the $E_{\rm p}$-$E_{\rm iso}$ diagram (\S \ref{sec:compar study}) and showed that the photospheric model best explains the burst. Our conclusions are further examined by a direct physical model fit (\S \ref{sec:phys mo}) and supported by a host galaxy candidate with appropriate redshift found in the 
location error box (\S \ref{sec:host galaxy}). A brief summary is presented in \S \ref{sec:summary}.

\section{Data Reduction and Analysis} \label{sec:data_ana}

GRB 210121A was detected by several other larger missions, including \fermi, HXMT. \fermi was launched in 2008, which comprises two scientific instruments, the Large Area Telescope \citep[LAT;][]{2009ApJ...697.1071A} and the Gamma-Ray Burst Monitor \citep[GBM;][]{2009ApJ...702..791M} and covers a broad energy band from 8 keV to $\sim 100$ GeV. 
The Hard X-ray Modulation Telescope \citep[HXMT;][]{GCN.29346},\textit{Insight}-HXMT, China’s first X-ray astronomy satellite, was launched in 2017 and consists of three main payloads, namely the high energy X-ray telescope (HE, 20-250 keV), the medium energy X-ray telescope (ME, 5-30 keV) and the low energy X-ray telescope (LE, 1-15 keV) \citep{Li:2007NuPhS,2018SPIE10699E..1UZ,Zhang:2020SCPMA}. {However, the CsI detectors of HE can monitor the $>$100 keV gamma-ray sky unocculted by the Earth. The measured energy range of CsI is 40-600 keV for Normal Gain mode and 200-3000 keV for low-gain mode \citep{2020JHEAp..27....1L}}. This study unitizes the high-energy data from all those telescopes as well as the GRID and GECAM missions aforementioned.

\subsection{Light Curves} \label{sec:lighcurves}

\begin{figure}
 \label{fig:light_curve}
 \includegraphics[width=0.47\textwidth]{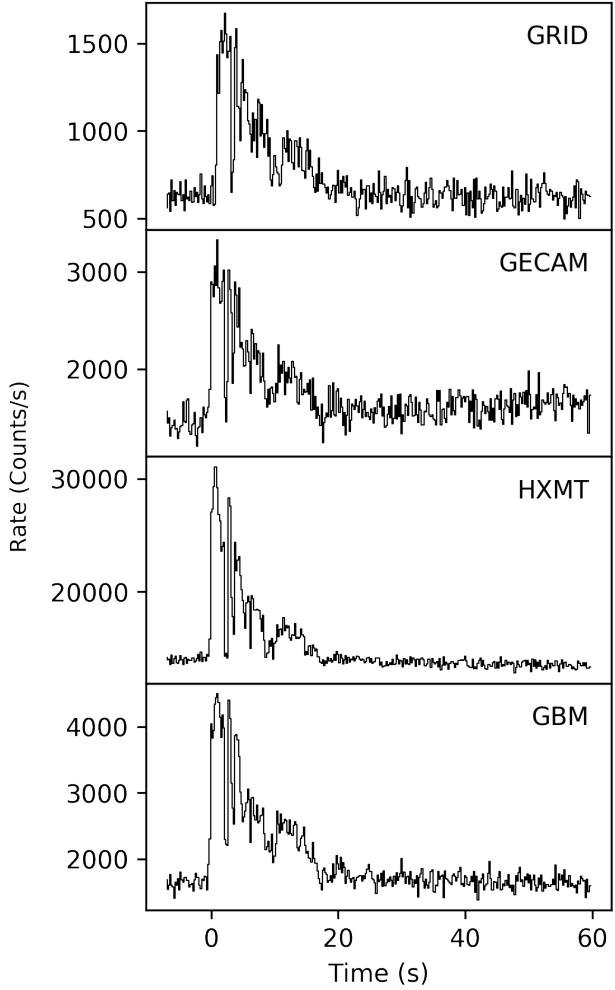}
 \caption{Light curves of the four missions. The bin size is set to 0.2 s for all light curves. The top panel shows the light curve within the energy range from 30 keV to 2000 keV by combining the data from all four GRID detector units. The second panel shows the light curve of detector B01 of GECAM. The third panel shows the light curve of the combined 18 CsI detectors of HXMT/HE in low-gain mode. The bottom panel shows the light curve of the NaI detector n3 of \fermi/GBM.}
\end{figure}

GRB 210121A almost simultaneously triggered HXMT \citep[2021-01-21T18:41:48.750 UTC,][]{2021GCN.29346....1X} and GECAM \citep[2021-01-21T18:41:48.800 UTC,][]{2021GCN.29347....1P}. For simplicity, we take a unique T$_0$ = 2021-01-21T18:41:48.800 UTC and align all the data accordingly. 
The four-mission light curves, which are all binned with 0.2 s, barycenter corrected, and aligned to GECAM trigger time,  are plotted together in Figure \ref{fig:light_curve}. The light curves are fully consistent with each other in the finest details, confirming the validation of the data of the four missions. 

\begin{figure}
\centering
\includegraphics[width=0.47\textwidth]{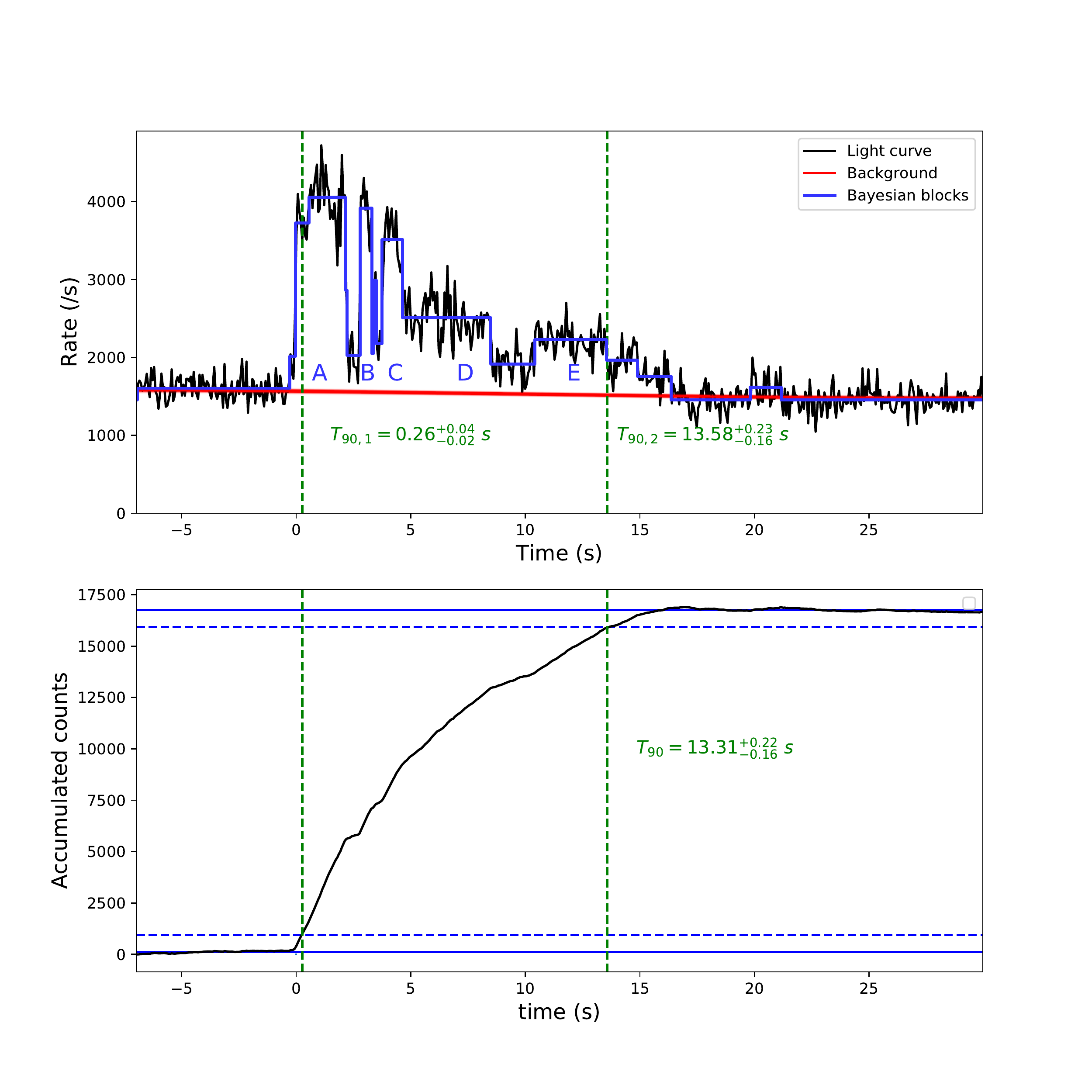} 
\label{fig:T_90}
\caption{The $T_{90}$ calculation. The upper panel shows the light curve (black line) and Bayesian blocks (blue line) derived from the NaI detector n3 of Fermi-GBM in the energy range from 8 keV to 800 kev}. The red line represents the background level. In the lower panel, the black line shows the accumulated counts. The blue solid (dashed) lines are drawn at the 0\% (5\%) and 100\% (95\%) of the accumulated counts. The vertical green dashed lines represent the $T_{90}$ in both panels.
\end{figure}

Following \cite{Y.S.Yang2020ApJ} and \cite{JunYang2020ApJ}, we calculate the burst duration, $T_{90}$, in 8 keV - 800 keV energy band using the continuous time-tagged event (CTTE) data of the \fermi/GBM detector n0 with bin size = 0.064s. The results are shown in Figure \ref{fig:T_90}. With a $T_{90}$ value of $13.31_{-0.16}^{+0.22}$ s (also see Table \ref{tab:para}), GRB 210121A definitely belongs to the long GRB population. 

We notice that several substructures clearly present in the light curve. Guided by the Bayesian block methodology \citep[blue histogram in the upper panel of Figure \ref{fig:T_90};][]{Scargle2013ApJ}, we divided the burst into five main emission episodes at 0.05 effective significant level, as listed in Table \ref{tab:spec_para}, and performed spectral analysis for each of them in \S \ref{sec:spectal_analysis}.

\begin{table}
\begin{center}
\label{tab:para}
\caption{Observational properties of GRB 210121A}
\begin{tabular}{ll}
\hline
\hline
Observed Properties & GRB 210121A \\
\hline
$T_{\rm 90}$ (sharp peak only) (s) & $13.31_{-0.16}^{+0.22}$\footnote{All errors correspond to the 1-$\sigma$ credible intervals.}\\
Total duration (s) & $\sim16.31$ \\
$\alpha$ at peak & $-0.52_{-0.05}^{+0.06}$ \\
$E_{\rm p}$ at peak (keV)& $1274.50_{-118.34}^{+192.64}$ \\
Time-integrated $\alpha$\footnote{The time-integrated spectral parameters are measured over the total duration.} &$-0.59_{-0.02}^{+0.02}$ \\
Time-integrated $E_{\rm p}$ (keV) & $954.33_{-38.62}^{+42.41}$\\
Total fluence\footnote{The total fluence and peak flux are calculated in the 10 keV$-$10 MeV energy band.} ($10^{-4}\ \rm erg\ cm^{-2}$) & $1.23_{-0.07}^{+0.08}$ \\
Peak flux ($10^{-5}\ \rm erg\ cm^{-2}\ s^{-1}$)&$2.19_{-0.36}^{+0.48}$ \\
z inferred by $E_{\rm p}-E_{\rm\gamma,iso}$ relation & $0.3-3.0$ \\
\multirow{2}{*}{Nearest host galaxy candidate} & J010725.95-461928.8\\
& (z $\sim$ 0.319)\\
\bottomrule
\end{tabular}
\end{center}
\leftline{\bf{Notes.}}
\end{table}

\subsection{Spectral Lags} \label{sec:lag}

Spectral lag, attributed to the fact higher-energy gamma-ray photons always arrive earlier than the lower-energy gamma-ray photons, is a common phenomenon in GRBs during their prompt emission epochs \citep{Norris86ApJ,Norris00ApJ, Band97ApJ,Chen05ApJ}. 
Several physical models, such as the curvature effect of relativistic jet and rapidly expanding spherical shell, have been proposed to explain the spectral lags \citep{Ioka01ApJL,Shen05MNRAS,Lu06MNRAS,Shenoy13ApJ,Uhm&Zhang16}. Statistically speaking, long GRBs are always characterized by obvious spectral lags, whereas lags of short GRBs are always negligible \citep{norris96,yi06}.

\begin{figure}
\begin{tabular}{l}
 \includegraphics[width=0.96\linewidth]{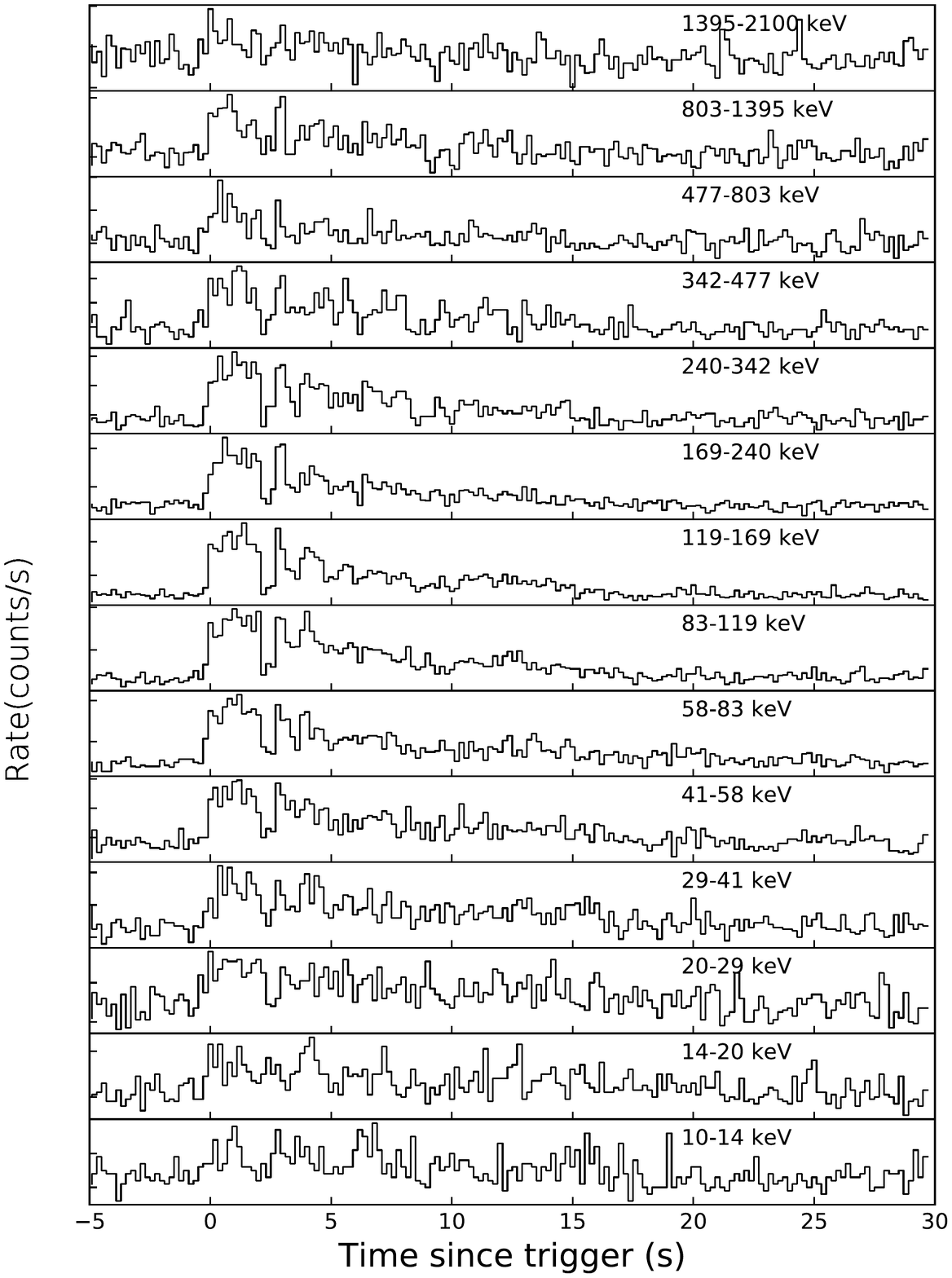}
 \\
 \includegraphics[width=0.96\linewidth]{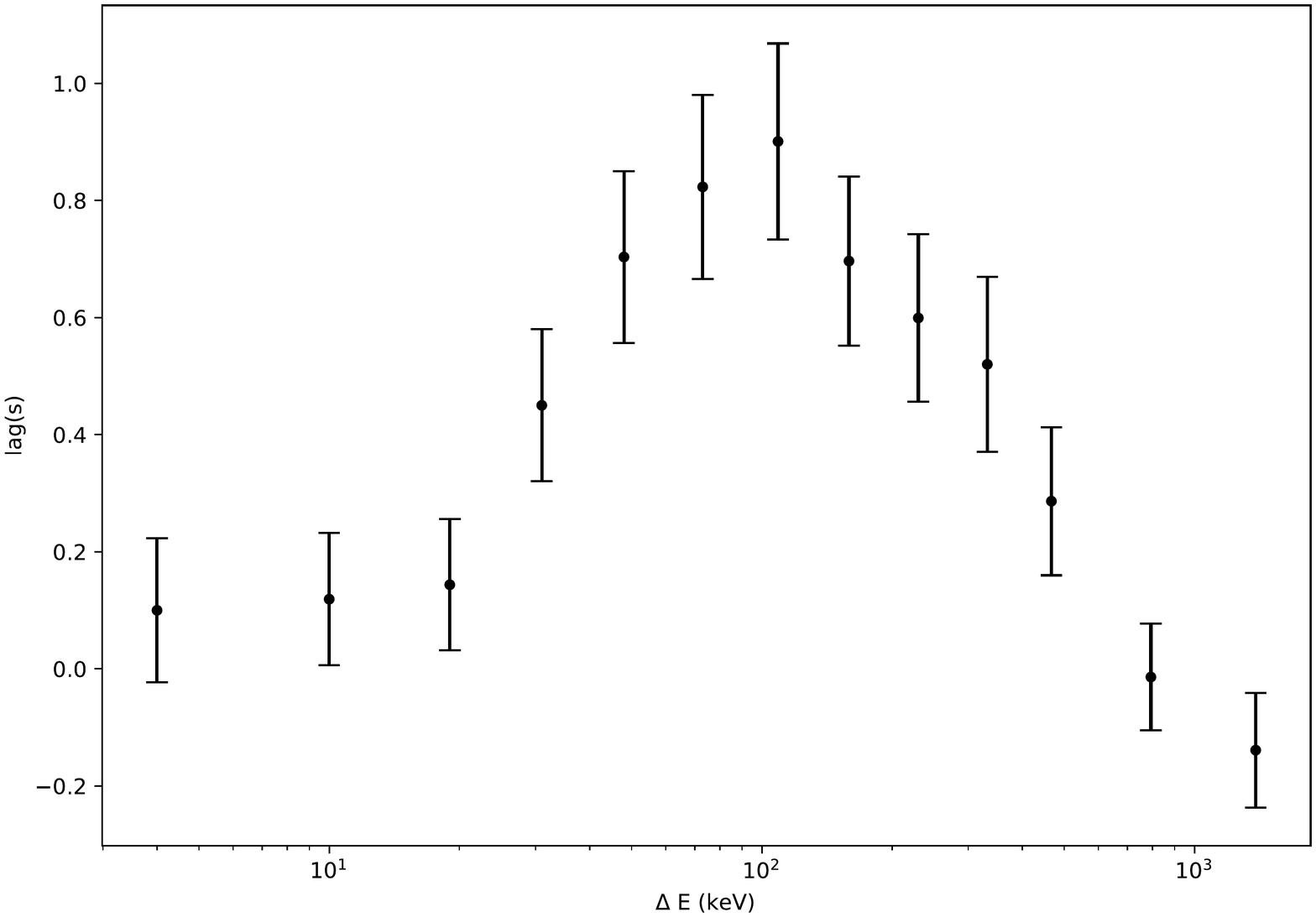}
\end{tabular}
\label{fig:lags}
\caption{Multi-wavelength light curves and spectral lag calculation. {\it Top}: the multi-wavelength light curves derived from the Fermi GBM-n0 and GBM-b0. {\it Bottom}: energy-dependent spectral lags between the lowest energy band (10-14 keV) and any higher energy bands. The error bars represent 1-$\sigma$ uncertainties.}
\end{figure}

To calculate the spectral lags, we first extract the light curves in different energy ranges from \textit{Fermi}/GBM NaI detector n0 and BGO detector b0. The multi-wavelength light curves are shown in Figure \ref{fig:lags} (top panel). After selecting the main emission range of 0-12 s, we calculate the lags between the lowest energy and any other bands following the method in \citet{2012ApJ...748..132Z}. The results are shown in Figure \ref{fig:lags} (bottom panel). A turn-over presents at the lag v.s. $\Delta E $ plot, which has noticed in some other long GRBs \citep[e.g.,][]{Wei17,Du21}.

\subsection{Spectral Analysis}\label{sec:spectal_analysis}

Both time-integrated and time-resolved spectral analyses are performed over the whole burst period, from -0.01 to 16.3 s. The five time-dependent slices, A to E\footnote{We note that there is an apparent dip after slice A. Including such a dip, however, does not significantly affect the spectral fitting results, as listed in Table \ref{tab:spec_para}}, are obtained by the Bayesian block method as mentioned in \S \ref{sec:lighcurves} and illustrated in Figure \ref{fig:T_90}. The photon flux in slice A is so bright that we can further divide them into five slices (A1-A5), each containing enough photon counts for spectral fitting. The boundaries of each slice are listed in Table \ref{tab:spec_para}.

\begin{table*}\label{tab:spec_para}
\caption{The spectral fitting results of GRB 210121A}
\setlength{\tabcolsep}{0.8pt}
\renewcommand\arraystretch{1.5}
\centering
\begin{tabular}{c|cc|ccccc|cccccc}
\hline
\hline
\multirow{4}{*}{ID}&\multirow{4}{*}{$t_{\rm start}$}&\multirow{4}{*}{$t_{\rm stop}$} & \multicolumn{5}{c|}{CPL} & \multicolumn{6}{c}{mBB}\\
\cline{4-14}
 &&& Flux & $\alpha$ & $E_{p}$ & $\dfrac{\rm pgstat}{\rm dof}$ & BIC & Flux & $m$ & $kT_{\rm min}$ & $kT_{\rm max}$ & $\dfrac{\rm pgstat}{\rm dof}$ & BIC\\
 &&& ($10^{-6}\ {\rm erg}$ & & (keV) & & &($10^{-6}\ {\rm erg}$& & (keV) & (keV) & & \\
 &&& ${\rm cm^{-2}\ s^{-1}}$) & & & & &${\rm cm^{-2}\ s^{-1}}$)& & & & & \\
 
\hline
A1&$-0.01$& 0.43 & $21.9_{-3.6}^{+4.8}$ &$-0.52_{-0.05}^{+0.06}$ & $1274.5_{-118.3}^{+192.6}$ & $\dfrac{294.6}{351}$ & $312.16$ & $22.5_{-5.0}^{+4.6}$& $0.21_{-0.11}^{+0.07}$ &$8.51_{-2.25}^{+4.36}$ &$636.60_{-90.27}^{+109.69}$ & $\dfrac{296.5}{350}$ & $320.00$ \\
A2& 0.43& 0.87 & $24.4_{-2.8}^{+3.8}$ &$-0.22_{-0.07}^{+0.07}$ & $1040.3_{-74.4}^{+98.1}$ & $\dfrac{316.0}{351}$ & $333.56$ & $23.4_{-3.3}^{+4.2}$&$0.67_{-0.09}^{+0.01}$ &$4.48_{-0.08}^{+8.30}$ & $401.66_{-17.64}^{+76.74}$ & $\dfrac{316.6}{350}$ & $340.10$ \\ 
A3&0.87& 1.32 & $21.5_{-2.7}^{+3.1}$ &$-0.33_{-0.05}^{+0.07}$ & $880.1_{-73.8}^{+85.1}$ & $\dfrac{348.9}{351}$ & $366.52$ & $20.9_{-2.7}^{+3.6}$ &$0.42_{-0.21}^{+0.05}$ &$7.88_{-1.51}^{+6.94}$ & $384.38_{-26.07}^{+79.34}$& $\dfrac{352.1}{351}$ & $375.57$\\
A4&1.32& 1.76 & $13.1_{-1.5}^{+2.1}$ &$-0.25_{-0.08}^{+0.07}$ & $627.3_{-42.9}^{+69.7}$ & $\dfrac{317.7}{351}$ & $335.35$& $14.2_{-2.3}^{+3.6}$ &$0.11_{-0.36}^{+0.14}$&$16.97_{-3.08}^{+9.32}$ & $357.37_{-46.06}^{+133.04}$& $\dfrac{317.7}{350}$ & $341.18$\\
A5&1.76& 2.20 & $10.8_{-1.5}^{+2.1}$ &$-0.24_{-0.10}^{+0.10}$ & $581.1_{-55.3}^{+76.1}$ & $\dfrac{305.2}{351}$ & $322.80$ & $19.7_{-9.7}^{+4.8}$ &$0.47_{-0.09}^{+0.19}$&$27.89_{-5.52}^{+3.49}$ & $1085.58_{-443.50}^{+179.05}$ & $\dfrac{301.1}{350}$ & $324.60$\\
\hline
A&$-0.01$& 2.20 & $18.9_{-1.3}^{+1.3}$ &$-0.38_{-0.03}^{+0.03}$ & $921.7_{-43.2}^{+41.8}$ & $\dfrac{449.7}{351}$ & $467.33$ & $18.9_{-1.4}^{+1.8}$& $0.28_{-0.08}^{+0.06}$ &$10.98_{-1.76}^{+2.47}$ & $442.96_{-28.59}^{+42.82}$ & $\dfrac{459.9}{350}$ & $483.36$\\

A+dip&-0.01 & 2.80 & $15.6_{-1.1}^{+1.3}$ &$-0.43_{-0.03}^{+0.03}$ &$950.4_{-43.67}^{+56.70}$ &$\dfrac{447.5}{351}$ &465.15 &$15.4_{-1.3}^{+1.6}$ &$0.29_{-0.07}^{+0.05}$ &$8.62_{-1.68}^{+2.17}$ &$438.45_{-27.42}^{+39.98}$ &$\dfrac{462.5}{350}$ &458.62
\\
\hline

B&2.80& 3.30 & $19.1_{-2.7}^{+3.6}$ & $-0.42_{-0.07}^{+0.05}$ & $960.7_{-80.8}^{+134.7}$ & $\dfrac{293.7}{351}$ & $311.33$ & $23.2_{-4.8}^{+5.7}$& $-0.14_{-0.12}^{+0.10}$ & $22.29_{-3.51}^{+4.24}$ & $752.89_{-116.45}^{+178.98}$ & $\dfrac{283.1}{350}$ & $306.61$\\

C&3.80& 4.60 & $13.4_{-1.9}^{+2.3}$ &$-0.54_{-0.05}^{+0.06}$ & $905.2_{-85.0}^{+117.4}$ & $\dfrac{331.9}{351}$ & $349.50$ & $12.3_{-1.5}^{+2.5}$& $0.31_{-0.12}^{+0.01}$ &$3.47_{-0.43}^{+3.28}$ & $366.94_{-15.41}^{+76.85}$ & $\dfrac{336.9}{350}$ & $360.37$ \\
D&4.60& 10.40 & $5.6_{-0.6}^{+1.1}$ &$-0.65_{-0.05}^{+0.03}$ & $941.9_{-51.2}^{+147.9}$ & $\dfrac{599.8}{351}$ & $617.44$ & $5.0_{-0.5}^{+0.8}$& $0.21_{-0.08}^{+0.01}$ &$2.99_{-0.06}^{+2.42}$ & $375.28_{-16.80}^{+55.96}$ & $\dfrac{609.4}{350}$ & $632.86$ \\
E&10.40& 16.30 & $3.8_{-0.6}^{+0.9}$ &$-0.76_{-0.05}^{+0.04}$ & $960.9_{-100.3}^{+186.5}$ & $\dfrac{501.2}{351}$ & $518.84$ & $3.2_{-0.5}^{+0.7}$ & $0.13_{-0.08}^{+0.01}$ &$2.20_{-0.25}^{+1.88}$ & $362.27_{-22.54}^{+67.79}$& $\dfrac{503.7}{350}$ & $527.23$\\
\hline
&$-0.01$& 16.30 & $7.5_{-0.4}^{+0.5}$ &$-0.59_{-0.02}^{+0.02}$ & $954.3_{-38.6}^{+42.4}$ & $\dfrac{1104.2}{351}$ & $1121.80$ & $6.7_{-0.4}^{+0.4}$ &$0.27_{-0.03}^{+0.02}$& $3.10_{-0.61}^{+1.21}$ & $379.82_{-11.84}^{+20.98}$& $\dfrac{1148.7}{350}$ & $1172.17$\\
\bottomrule

\end{tabular}
\end{table*}

\begin{figure*}
\begin{tabular}{lll}
\includegraphics[angle=0,width=0.3\textwidth]{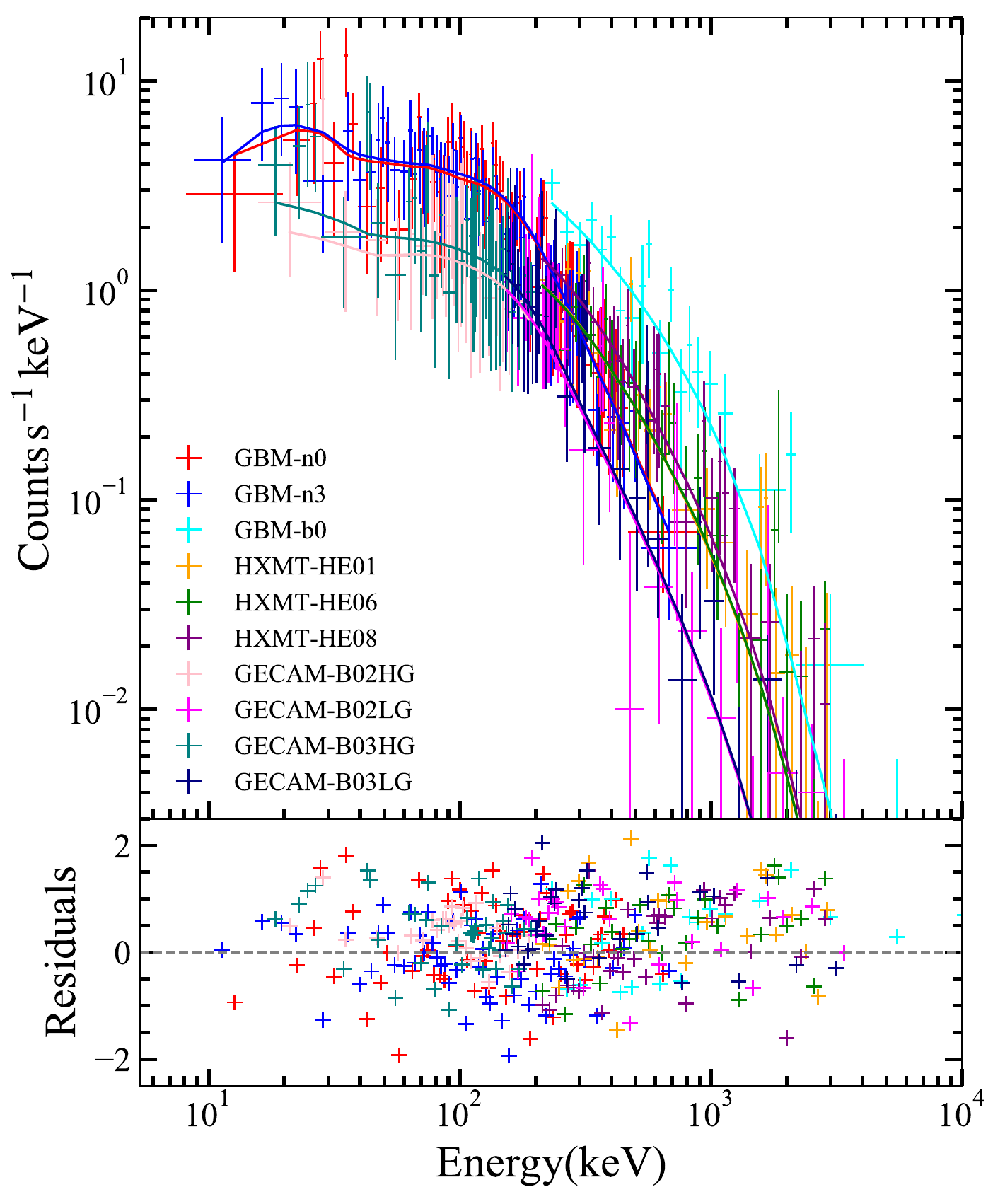} &
\includegraphics[angle=0,width=0.3\textwidth]{figure4B.pdf} &
\includegraphics[angle=0,width=0.3\textwidth]{figure4C.pdf}
\end{tabular}
\caption{The joint spectral fitting of multi-mission energy spectra in the time interval from 4.5 s to 6.0 s. The left panel shows the observed photon count spectra (points) and convolved photon spectra (lines). Due to the unavailability of the response files, GRID spectra are not involved in the joint fit. The middle panel shows the deconvolved photon spectrum. The right panel is the corner diagram, in which the red plus signs represent the best-fit value and the 1-$\sigma$ credible intervals. The histogram and the contours show the 1D probability distributions and the 2D likelihood maps, respectively.}
\label{fig:BB_corner}
\end{figure*}

For each slice, we extract the corresponding source and background spectra, and the corresponding instrumental response files following the procedures described in \cite{2018NatAs...2...69Z}. Since the energy range of \fermi/GBM is the broadest one among the four missions, for simplicity, we only employ the GBM data in our spectral analysis. The spectral data are obtained from three detectors with relatively small viewing angles (i.e. the NaI detector n0 and n3, as well as the BGO detector b0). Nevertheless, we have confirmed that the joint-mission spectral fitting (e.g., in Figure \ref{fig:BB_corner}) using all four missions yields consistent results to those with GBM data only.

For each slice, we perform detailed spectral fit using our self-developed Monte-Carlo fitting tool \textit{McSpecFit} \citep{2016ApJ...816...72Z, 2018NatAs...2...69Z} with four frequently used empirical models, namely the band function (Band), the blackbody (BB), the multi-color blackbody (mBB)\footnote{The mBB model is defined as $N(E)=8.0525(m+1)K[(T_{\rm max}/T_{\rm min})^{m+1}-1]^{-1}(kT_{\rm min}/{\rm keV})^{-2} I(E)$ \citep{2018ApJ...866...13H}, where $ I(E)=[E/(kT_{\rm min})]^{\rm m-1} \int_{E/(kT_{\rm max)}}^{E/(kT_{\rm min})} x^{\rm 2-m}/(e^x-1)dx$, x = $E/(kT)$, $K$ = $L_{39}$/$D_{L, 10 kpc}^2$, and m is the power-law index of distribution. The temperature of the blackbody is from $T_{\rm min}$ to $T_{\rm max}$.}, 
the single power law (PL), and the cutoff power law (CPL) models\footnote{The CPL model can be expressed as $
 N(E)=A{E}^{\alpha}{\rm exp}[-(\alpha+2)E/E_{\rm p}] 
 \label{N(E)}
$ \citep{2016A&A...588A.135Y},
where $\alpha$, $A$ and $E_{\rm p}$ is the photon index, normalized coefficient, and peak energy, respectively.}. The ratio of Profile-Gaussian likelihood to the degree of freedom \citep[PGSTAT/dof,][]{1996ASPC..101...17A} and the Bayesian information criterion \citep[BIC,][]{1978AnSta...6..461S} are taken into account to test the goodness-of-fit.

\begin{figure}
 \label{fig:spectral_evolution_CPL}
 \centering
 \includegraphics[width=0.47\textwidth]{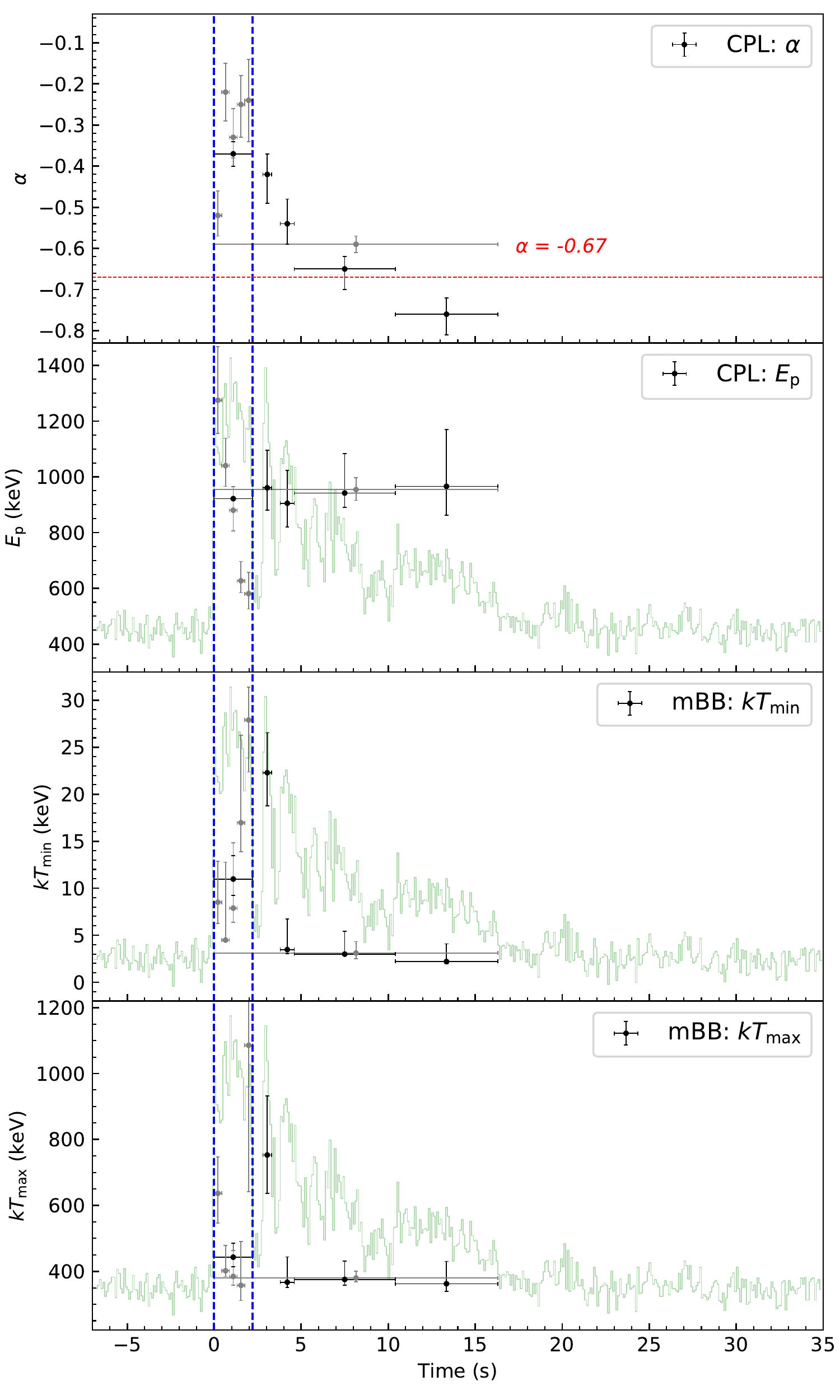} 
 \caption{The spectral evolution of CPL model and mBB model. The horizontal errors represent the time spans, and the vertical errors indicate the 1-$\sigma$ uncertainties of the best-fit parameters.
 In the first panel, the red horizontal dashed line represents the synchrotron death line. The blue vertical dashed lines mark slice A.}
\end{figure}

Our results show that the CPL model is the preferred one for all the time-resolved slices. The best-fit parameters obtained by the CPL models are listed in Table \ref{tab:spec_para}. The corresponding spectral evolution is plotted in Figure \ref{fig:spectral_evolution_CPL}. The peak energy constrained by the CPL model is typically around 1 MeV throughout the burst and exhibits strong spectral evolution from slices A1 to A5. Some previous studies on multi-pulse long GRBs \citep[e.g.,][]{2012ApJ...756..112L} found that the $E_{\rm p}$ evolution displays two prevailing trends: (1) hard-to-soft then tracking, meaning $E_p$ first shows a hard-to-soft evolution at the beginning of the burst then tracks the intensity level; (2) the intensity-tracking, meaning $E_{\rm p}$ tracking the intensity all the time during a burst. However, the $E_{\rm p}$ evolution of GRB 210121A is different from neither of the two above. The $E_{\rm p}$ follows the hard-to-soft pattern throughout the first pulse in slices A1-A5, and remains a high value till the final stage of the burst. The best-fit low-energy photon index, $\alpha$, evolves rapidly and tracks the flux level. Moreover, $\alpha$ systematically exceeds the synchrotron ``death line" \citep{1998ApJ...506L..23P} defined by $\alpha=-2/3$ and reaches the highest value of $\sim -0.2$ in several slices, which indicates that the spectra are thermal-like \citep{2019ApJ...882...26M}.

The hard low-energy photon index motivates us to re-evaluate the fit using thermal-like spectral models. Although the single black body model was unacceptable in the time-resolved spectral fit, we found the multi-color black body model (mBB), on the other hand, can well explain the time-resolved spectra with adequate goodness of fit (Table \ref{tab:spec_para}). The corresponding best fit values of $kT_{min}$ and $kT_{max}$ are also plotted in Figure \ref{fig:spectral_evolution_CPL}.

\section{Comparison Study} \label{sec:compar study}

\subsection{Comparison With Other Long GRBs} \label{sec:compar}

\begin{figure}
 \label{fig:probability}
 \centering
 \includegraphics[width=0.47\textwidth]{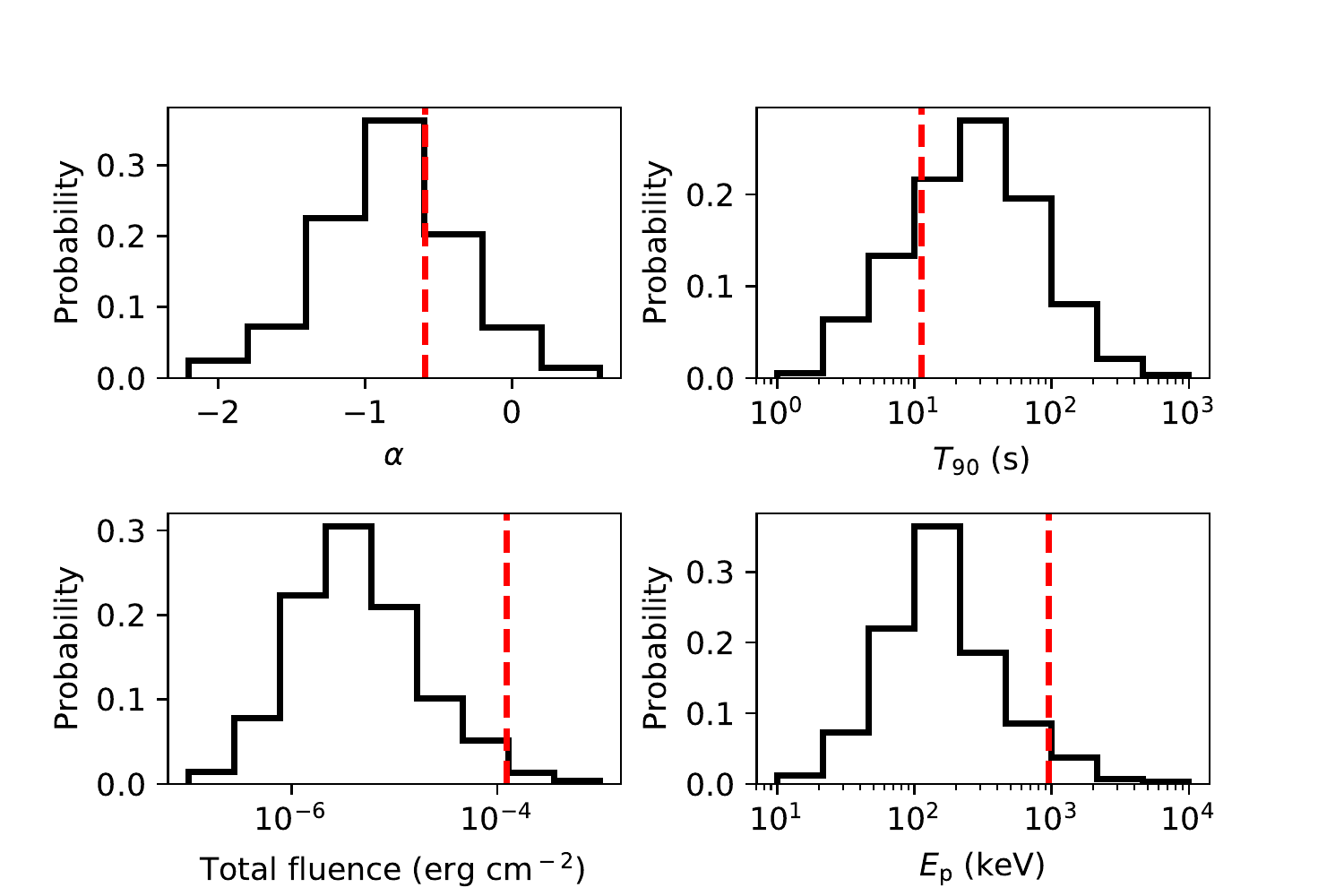} 
 \caption{Probability distributions of some characteristic properties of long GRBs. The red vertical dashed lines highlight GRB 210121A. The long GRB sample is from the Fermi/GBM catalog \citep{Gruber_2014, von_Kienlin_2014, Bhat_2016, von_Kienlin_2020}.}
\end{figure}

Based on above data analyses on GRB 210121A, we have obtained its temporal characteristic parameters such as $T_{90}$ of 11.78 s and some spectral properties including the spectral index $\alpha$, the time-integrated peak energy being $954.33_{-38.62}^{+42.41}$ keV and the total fluence of $1.23^{+0.08}_{-0.07}\times10^{-4}\,\rm{erg\,cm^{-2}}$ calculated within 10-10,000 keV energy range. The above properties are also listed in Table \ref{tab:para}. To check how special GRB 210121A is, we mark the above values as red dashed lines (see Figure \ref{fig:probability}) within the distributions of some characteristic parameters of long GRBs from the Fermi/GBM burst catalog \citep{Gruber_2014, von_Kienlin_2014, Bhat_2016, von_Kienlin_2020}. As shown in Figure \ref{fig:probability}, GRB 210121A is relatively special compared to the majority of the bursts due to its significantly high peak energy and fluence in the whole long GRB sample.

\subsection{Placement On $E_{\rm p}$-$E_{\rm\gamma, iso}$ Correlation} \label{sec:Ep-Eiso}

The $E_{\rm p}$-$E_{\rm\gamma, iso}$ relation \citep[a.k.a. Amati Relation;][]{Amati2002} has been used as a powerful tool to diagnose GRB classifications. As shown in Figure \ref{fig:Ep-Eiso}, typical long and short GRBs follow two separate tracks. In addition, a third track was recently found for those short GRBs originating from magnetar giant flares \citep[a.k.a. MGF-GRBs, e.g., GRB 200415A,][]{JunYang2020ApJ, Svinkin2021Natur, Roberts2021Natur,H.M.Zhang2020ApJ}. Since there is no redshift measurement for GRB 210121A, we assign its redshift as a free parameter ranging from $10^{-5}$ to 10 and overplot it with a dotted line in Figure \ref{fig:Ep-Eiso}. One can immediately infer from such a plot that the redshift of GRB 210121A should be within the range between 0.3 and 3.0 in order to be consistent with a long GRB. Other possibilities are almost ruled out as it is certainly not a short GRB nor a giant flare from a magnetar given the properties (e.g., non-negligible lags, long duration) presented in \S \ref{sec:data_ana}. In addition, the absence of a nearby host galaxy at redshift $\sim$ 0.0001 further rules out its possibility of being an MGF-GRB.

\subsection{Constraints From Photonsphere Death Line} 
\label{sec:deathline}

\begin{figure}
 \label{fig:Ep-Eiso}
 \centering
 \includegraphics[width=0.47\textwidth]{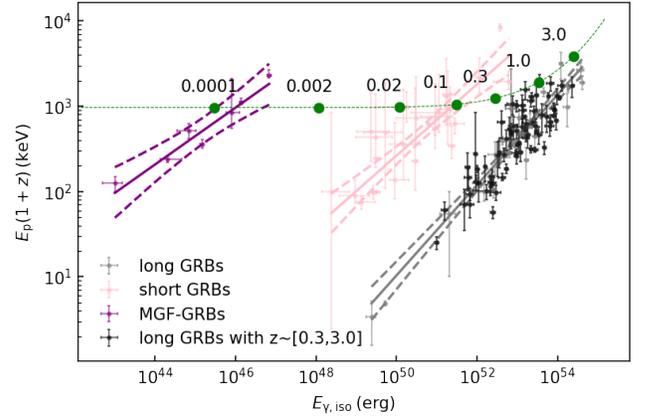}
 \caption{The $E_{\rm p, z}$ versus $E_{\rm\gamma,iso}$ correlation diagram. The black, pink, and purple solid lines are the best-fit correlations of the long, short, and MGF GRB samples. The green dashed line shows the trajectory of GRB 210121A by applying different redshift values, and the green points highlight some specific redshift values.}
\end{figure}

As shown in Figure \ref{fig:Ep-Eiso}, GRB 210121A is a significantly extreme case among those GRBs with a redshift between 0.3 and 3. The large $E_{\rm p}$ value of GRB 210121A may be subject to the so-called ``photonsphere death line" constraints as discussed in \cite{2012ApJ...758L..34Z}.

According to \cite{2012ApJ...758L..34Z}, the photosphere emission model predict an up-limit for the peak energy of a GRB in the form of

\begin{equation}
E_{\text{p}} \le \zeta kT_0 \simeq 1.2 \zeta L_{52}^{1/4}r_{0}^{-1/2} \rm{\ MeV} 
\label{eq:Ep}
\end{equation}
where $L_{52}$ is the luminosity in unit of $10^{52}$ erg $\rm s^{\rm -1}$, $r_0$ is the initial fireball radius in unit of $10^{7}$ cm, and the factor $\zeta$ is taken as 2.82 in this study which invokes a relativistic multi-color black body outflow. Accordingly, with a known $E_{\rm p}$ and $E_{\rm\gamma, iso}$ , one can calculate the lower limit of the initial fireball radius:

\begin{equation}
 r_{0} \le 1.44 \zeta^{2} \Bigl(\frac{E_{\rm p}}{\rm{MeV}} \Bigr)^{-2} \Bigl(\frac{E_{\rm\gamma, iso, 52}}{\Delta t}\Bigr)^{1/2} \times 10^7 \rm{\ cm}
\label{eq:r0r0} 
\end{equation}

By putting the observed $E_{\rm p} = 1274.5$ keV, $\Delta t = 0.44$ s and flux of $\sim 21.9\times 10^{-6}$ erg cm$^{-2}$ s $^{-1}$ into Equation (\ref{eq:r0r0}), we obtain an up-limit for $r_0$ of 


\begin{equation}
 r_{0} \le [3.10,\ 5.40]\times 10^7 {\rm cm}
\label{eq:r0r0r0} 
\end{equation}
for a redshift range of $z\sim $ [0.3,3.0]. Such an up-limit is fully consistent with the prediction of the standard fireball model \citep[e.g.,][]{2000ApJ...530..292M,2002ApJ...578..812M} as well as the mean acceleration radius ($\sim 10^8$ cm) of the fireball derived form observed data \citep{2015ApJ...813..127P}.

\section{Physical Models} \label{sec:phys mo}

\begin{figure*}
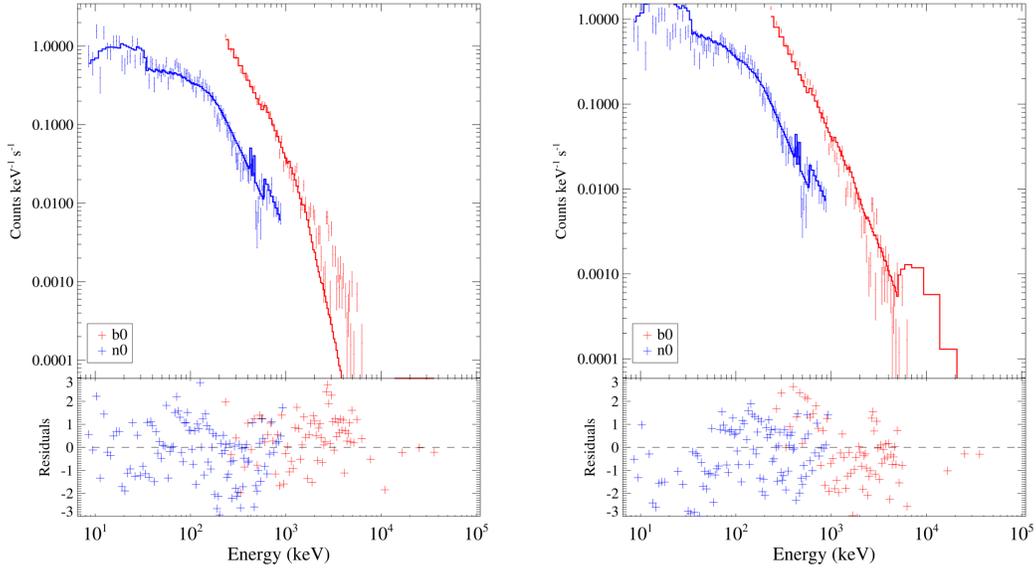

\centering
\subfigure{\includegraphics[width=0.38\textwidth]{figure8A.pdf}}
\hspace{0.01\linewidth}
\subfigure{\includegraphics[width=0.38\textwidth]{figure8B.pdf}}
\caption{The observed photon count spectra of the photosphere model and the synchrotron model.}
\label{fig:subfig}
\end{figure*}

\begin{figure*}
\centering
\subfigure{\includegraphics[width=0.38\textwidth]{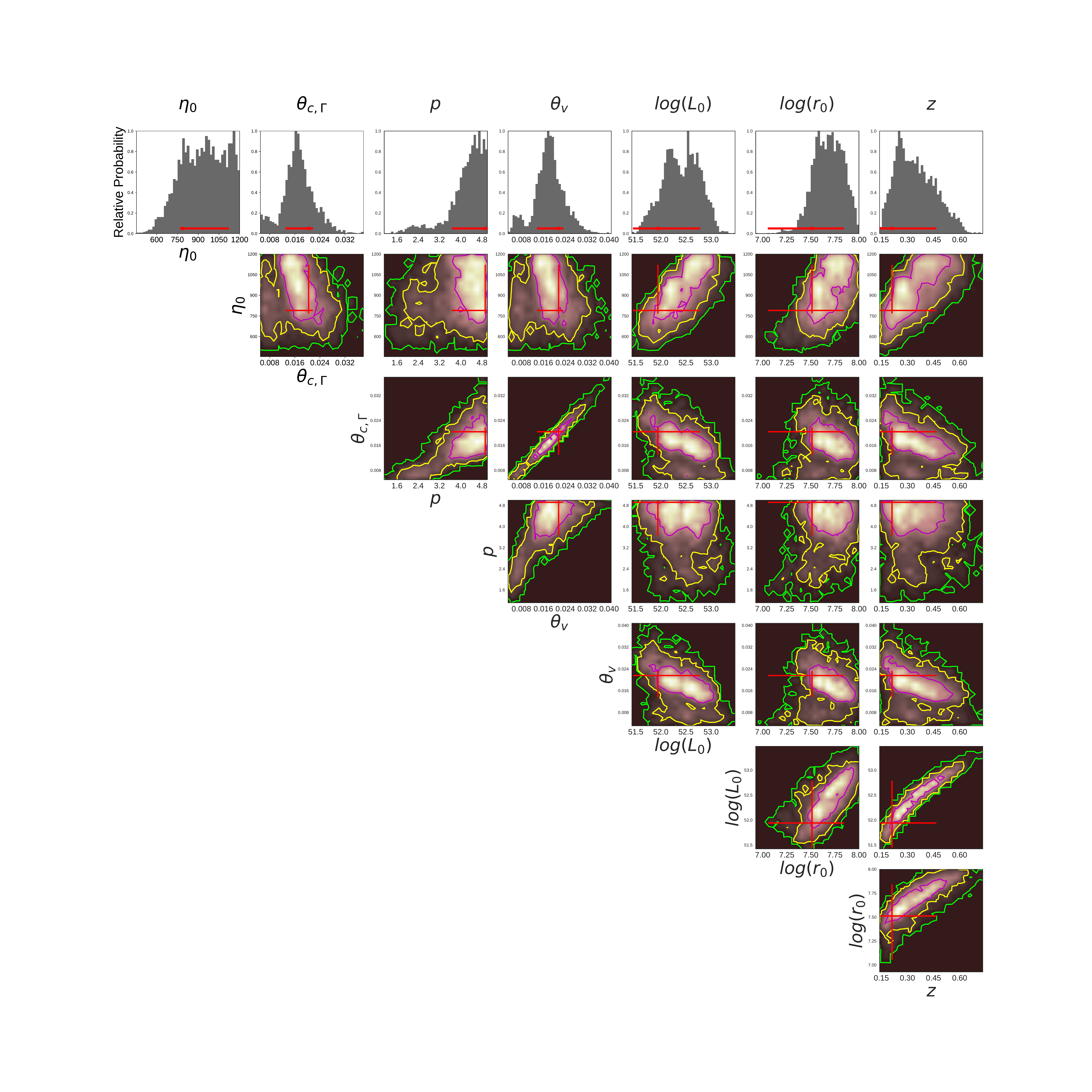}}
\hspace{0.01\linewidth}
\subfigure{\includegraphics[width=0.38\textwidth]{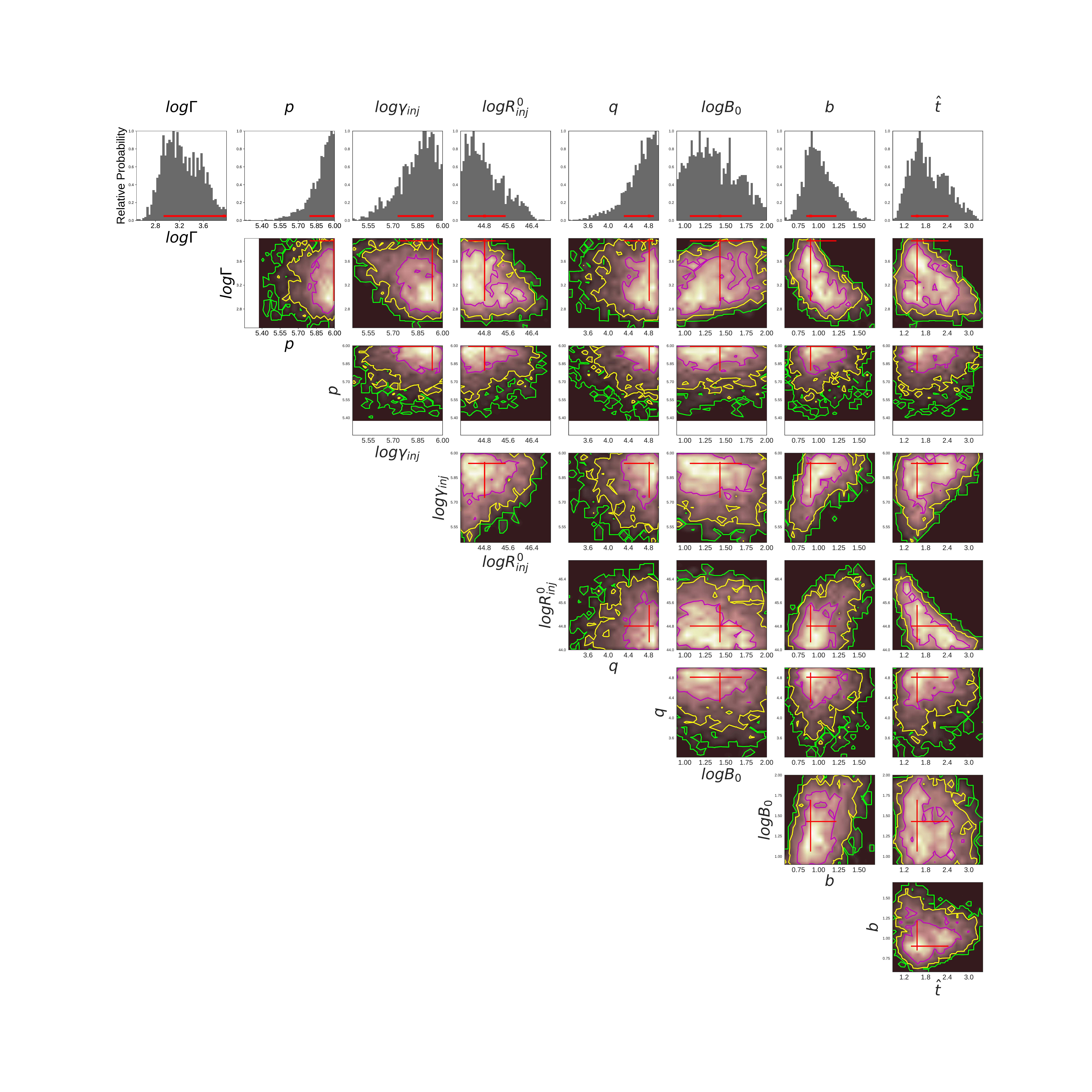}}
\caption{Parameter constraints of the photosphere model(left) and synchrotron model(right) fit for the spectrum in the time interval of slice A.}
\label{fig:fig_para}
\end{figure*}

This section employs two physical (photosphere v.s. synchrotron) radiation models and fits them directly to the same observed data. By comparing the goodness of fit and parameter constraints of the two fits, we can, from the first principle, further evaluate the radiation mechanisms that shape the observed spectra.

\subsection{Photosphere Model}
\label{sec:fit_photosphere}

\begin{table}
\centering
\renewcommand{\arraystretch}{1.5}
\caption{Fitting parameters of photosphere v.s. synchrotron model}
\begin{tabular}{cc|cc}
 \hline
 \hline
 \multicolumn{2}{c|}{Photosphere model}&\multicolumn{2}{c}{Synchrotron model}\\
 \hline
 Parameter & Range & Parameter & Range\\
 \hline
 $\eta_{0}$ & $789.82_{-22.13}^{+331.91}$ & log$\Gamma$ & $3.94_{-1.01}^{+0.04}$\\
 $p$ & $4.92_{-1.26}^{+0.08}$ & $p$ & $5.99_{-0.20}^{+0.00}$\\
 $\theta_{c,\Gamma}$ & $0.02_{-0.01}^{+0.001}$ & log$\gamma_{\rm inj}$ & $5.94_{-0.21}^{+0.01}$\\
 $\theta_{v}$ & $0.02_{-0.01}^{+0.002}$ & log$\left(\dfrac{R_{\rm inj}^{0}}{\rm s} \right)$ & $44.81_{-0.55}^{+0.71}$\\
 \multirow{2}{*}{log$\left(\dfrac{L_{0}}{{\rm erg} {\rm \ s}^{-1}}\right)$}& \multirow{2}{*}{$51.94_{-0.50}^{+0.84}$} & log$\left(\dfrac{B_{0}}{\rm G}\right)$&$1.43_{-0.37}^{+0.27}$\\
 & & $q$ & $4.81_{-0.50}^{+0.09}$\\
 log $\left(\dfrac{r_{\rm 0}}{{\rm cm}}\right)$& $7.51_{-0.46}^{+0.33}$ & $b$ & $0.90_{-0.05}^{+0.32}$\\
 $z$&$0.21_{-0.07}^{+0.25}$& $\hat{t}$ (s) & $1.56_{-0.17}^{+0.87}$\\
 BIC & 329.57 & BIC & 468.11\\
 $\dfrac{\rm PGSTAT}{\rm dof}$ & $\dfrac{291.1}{238}$ & $\dfrac{\rm PGSTAT}{\rm dof}$ & $\dfrac{424.1}{237}$\\
\bottomrule 
\end{tabular}	
\end{table}

We apply a structured jet photosphere model \citep{2013MNRAS.428.2430L, 2018ApJ...860...72M, 2021arXiv210704532M} to fit the first-pulse spectra. The flux density of this model (in unit of mJy) can be calculated
numerically in the form 
\begin{equation}
 F_{\nu}(E) = F_{\nu}(E;\eta_{0},p,\theta_{c,\Gamma},\theta_v,L_{0},r_{0},z)
\end{equation}
Seven parameters are involved, namely the baryon loading value $\eta_{0}$, the power-law decay index of for the baryon loading $p$, the angular width for the isotropic core of the baryon loading $\theta_{c,\Gamma}$, the viewing angle $\theta_{v}$, the outflow luminosity $L_{0}$, the initial radius of the fireball $r_{0}$, and the redshift $z$. As shown in Figure \ref{fig:subfig}, the model successfully fits the data with a PGSTAT/dof = 291.07/238.0 = 1.22. The best-fit parameters as well as their  constraints as shown in Figure \ref{fig:fig_para}. The best-fitting value of the initial raidus is $r_{0}$ = $3.2^{+3.7}_{-2.1}\times 10^{7}$ cm, which is consistent with the constraints in Eq. (3). 
Besides, the redshift is constrained to be $0.14 \le z \le 0.46$. 

The Lorentz factor at the photosphere can be discussed under two cases: saturated acceleration (Case I) and unsaturated acceleration (Case II).
According to our results, the Lorentz factor is calculated to be
\begin{equation}
\Gamma =
	\left\{
		\begin{array}{lll}
			\eta_{0} &= 789.8, & \text{Case I},\\
			\bigg[\frac{\sigma_{T}}{6m_{p}c}\frac{L(\theta)}{4\pi c^{2}\eta(\theta) r_0}\bigg]^{1/3} &= 405.4, & \text{Case II},
		\end{array}
	\right. 
\end{equation}
where $\eta(\theta) = \frac{\eta_{0}}{[(\theta/\theta_{c,\Gamma})^{2p}+1]^{1/2}}+1.2$ is the structured baryon loading parameter and $\theta$ is the angle measured from the symmetry axis of the jet. {Assuming the fireball is saturated accelerated and $\theta = 0$, the derived photosphere radius is $R_{ph} = \frac{1}{(1+\beta)\beta\eta^{2}(\theta)}\frac{\sigma_{T}}{m_{p}c}\frac{L(\theta)}{4\pi c^{2}\eta(\theta)}$ = 1.0 $\times 10^7$ cm. Compared with the saturated radius $R_{s} = r_0 \eta(\theta)$ = 2.6 $\times 10^7$ cm, the result of which is inconsistent with the condition that $R_s < R_{ph}$ in Case I. On the other hand, the Lorentz factor $\Gamma$ = 405.4 is obtained precisely in Case II, which is consistent with the average $\Gamma \sim 370$ \citep{2015ApJ...813..127P}}.

\subsection{Synchrotron Model} 
\label{sec:fit_synchrotron}

Similarly, we apply the synchrotron model \citep{2014NatPh..10..351U, 2016ApJ...816...72Z} to fit the same spectrum used in \S \ref{sec:fit_photosphere}. the redshift is assumed to be $z = 0.319$ based on the constraints in \S\ref{sec:Ep-Eiso}, \S \ref{sec:fit_photosphere} and \S\ref{sec:host galaxy}. The flux density of this model (in unit of mJy) is in the form 
\begin{equation}
 F_{\nu}(E) = F_{\nu}(E;\Gamma,p,\gamma_{inj},R_{inj}^{0},q,B_{0},b,\hat{t})
\end{equation}
Our fit can constrain eight parameters (Figures \ref{fig:subfig} and \ref{fig:fig_para}), including the Lorentz factor $\Gamma$, the power-law index of the electron spectrum $p$, the minimum Lorentz factor of electrons $\gamma_{inj}$, the normalized injection rate of electrons $R_{inj}^{0}$, the power-law index of the injection rate $q$, the initial magnetic filed $B_{0}$, the decaying factor of the magnetic field $b$ and the time at which electrons begin to radiate in the observer frame $\hat{t}$.

Compared to the photosphere model in \S \ref{sec:fit_photosphere}, the model is underfitting with a PGSTAT/dof $= 424.10/237.0 = 1.79$. Besides, some of the parameters and the derived parameters are unreasonable:
\begin{itemize}
 \item The bulk Lorentz factor in this model is as high as $10^{3.94}$, which is greater than the upper limit constrained in various ways \citep{2011ApJ...738..138R}.
 \item The photon index $\alpha = -(p-1)/2 = -2.5$. Not only is it inconsistent with the value in the CPL model, but also it is too soft compared the typical values of the the GRB sample \citep{2021arXiv210313528P}.
 \item The emission radius, $R_{0} = 2\Gamma^{2}c\hat{t} = 7.09\times 10^{18}$ cm, appears to be too large as a GRB emission radius.
\end{itemize}

\section{Host Galaxy Search} \label{sec:host galaxy}

\begin{figure}
 \centering
 \includegraphics[angle=0,width=0.47\textwidth]{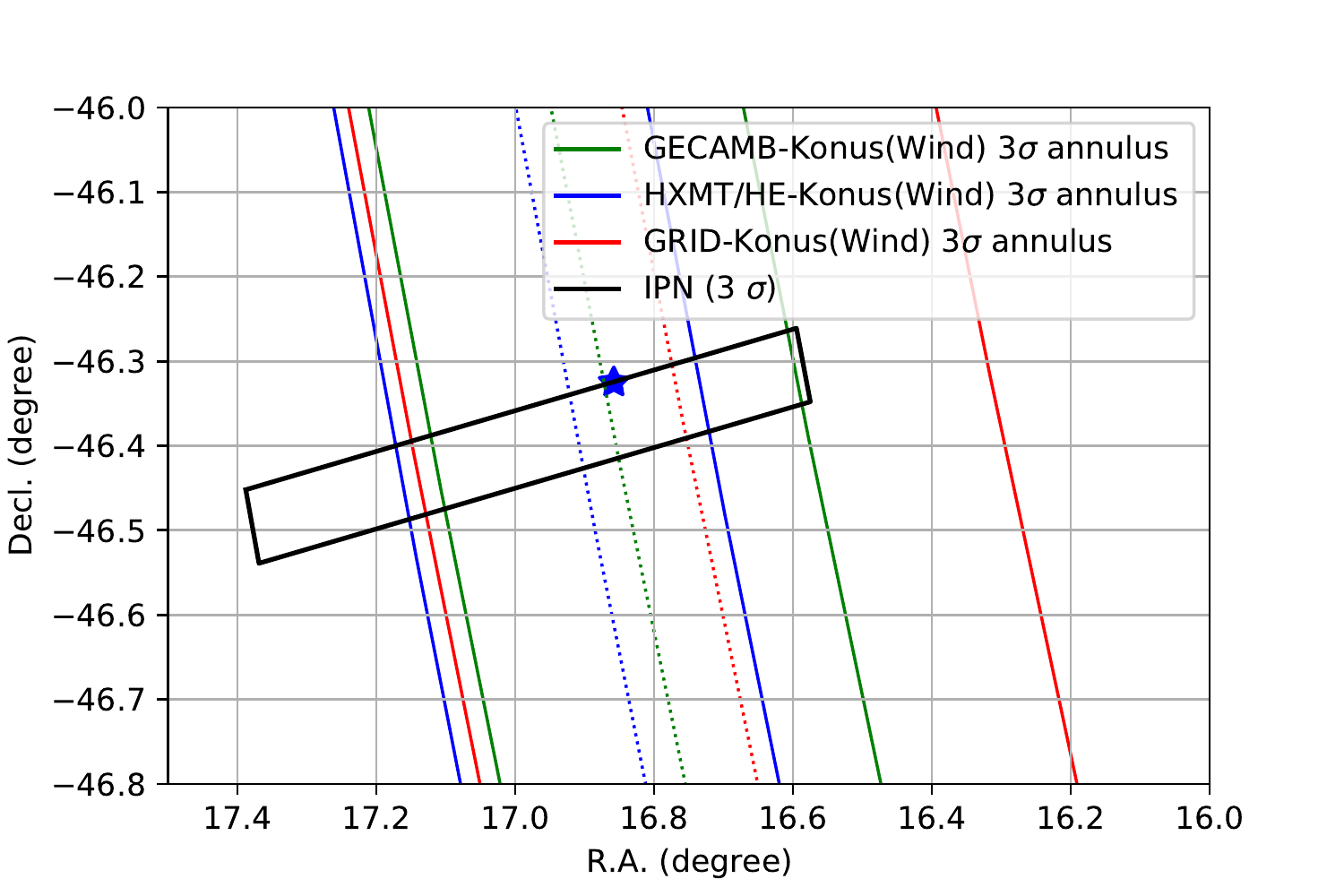}
 \caption{The localization of GRB 210121A. The black quadrangle is the IPN error box. The green, blue, and red solid lines are 3-$\sigma$ error lines of joint localization of GECAM-Konus (Wind), HXMT-Konus (Wind), and GRID-Konus (Wind) respectively. The host galaxy candidate, J010725.95-461928.8, is marked with a blue star.}
 \label{fig:loc}
\end{figure}

\begin{figure*}
 \centering
 \includegraphics[angle=0,width=1\textwidth]{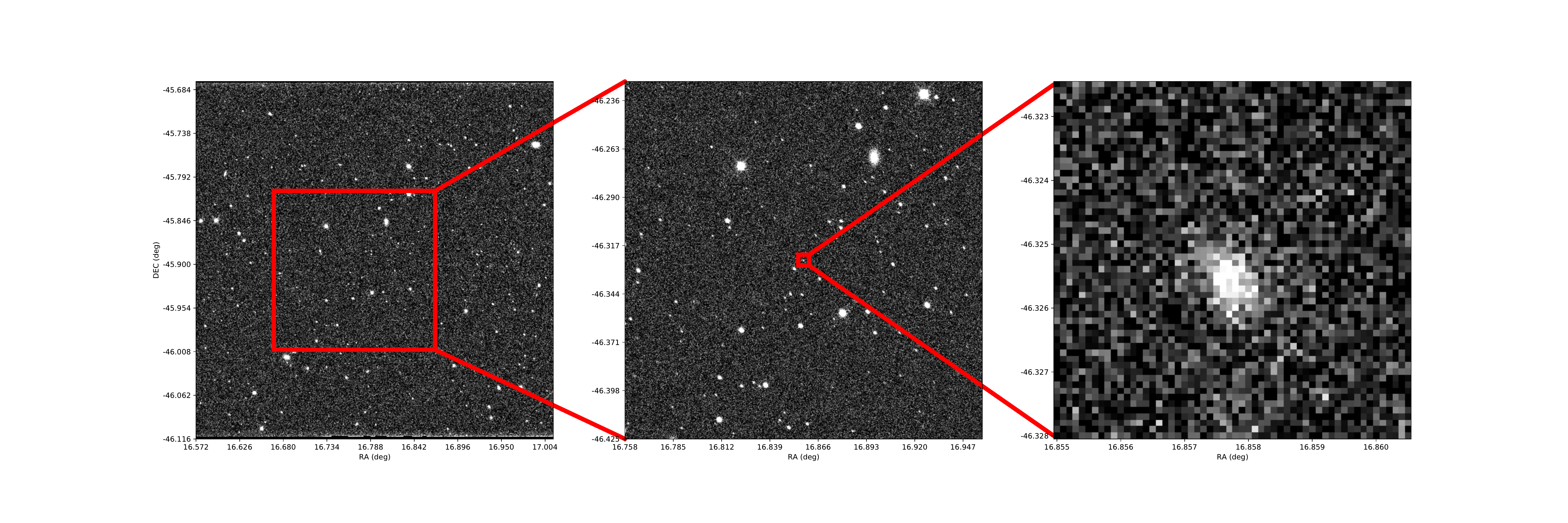}
 \caption{The follow-up observation on galaxy J010725.95-461928.8 with LCOGT. The left panel is the image of the area of sky near the localization of GRB 210121A. The middle panel is the area of sky localized in the error box constrained from Figure \ref{fig:loc}. The right panel is the host galaxy candidate J010725.95-461928.8.}
 \label{fig:gala}
\end{figure*}

As shown in \S \ref{sec:phys mo}, the physical photosphere model constrains the redshift to a range of [0.14, 0.46]. Combining the Amati relation requirements (i.e., z $\sim$ [0.30, 3.0])in \S \ref{sec:Ep-Eiso}, we finally narrow down the redshift of GRB 210121A to [0.30, 0.46].

We then search the field of GRB 210121A for its possible galaxies within the redshift range of [0.30, 0.46]. There is no optical and X-ray counterpart observed for this burst. To get the localization, {we make use of the Inter-Planetary Network (IPN) triangulation location of GRB 210121A, a 3-$\sigma$ error box of 181 arcmin$^2$ centered RA = 16.981 degree and DEC = -46.401 degree \citep{2021GCN.29348....1H}. Based on the public Konus-Wind data,} this IPN error box is further improved by involving the joint triangulations of GECAM-Konus (Wind), HXMT-Konus (Wind), and GRID-Konus (Wind), as shown in Figure \ref{fig:loc}, which provides the final location box for the host galaxy search.

We searched four catalogs, namely, the SIMBAD\footnote{\url{http://simbad.u-strasbg.fr/simbad/sim-fcoo}}, the NED\footnote{\url{http://ned.ipac.caltech.edu}}, the SuperCOSMOS\footnote{\url{http://ssa.roe.ac.uk/WISExSCOS.html}}, and the HyperLeda\footnote{\url{http://leda.univ-lyon1.fr}} within that error box. Our search yields only one galaxy, J010725.95-461928.8, in the SuperCOSMOS catalog, with RA $= 16.858$ degree and Dec $= -46.325$ degree and $z=0.319$ \citep{2016ApJS..225....5B}, which is likely the host galaxy of GRB 210121A. We followed up the galaxy with the Las Cumbres Observatory (LCOGT) at 2021-05-09T04:01:16.593 UTC. The resulted image in $R$ band is shown in Figure \ref{fig:gala}. The host galaxy candidate is clearly visible. However, no optical counterpart for this GRB was detected in our observation.

\section{Summary} \label{sec:summary}

After performing a comprehensive analysis of the high energy data of GRB 210121A and the host galaxy search, we can claim the burst originating from the photosphere emission at a typical fireball radius due to the following rationale:

\begin{enumerate}
\item The burst is characterized by a hard low-energy spectral index, likely due to thermal origin.\item To place the burst onto the long GRB track on Amati relation, the large values of $E_{\rm p}$ and fluence requires a redshift range of [0.3,3.0].
\item The physical photosphere model successfully fits the observed spectra and constrains the redshift in the range [0.14,0.46].
\item By overlapping 2 \& 3 , one can further constrain the redshift to a range of [0.30,0.46].
\item By searching the error box of the GRB field, we only find one galaxy within the redshift range of [0.30,0.46]. The galaxy is J010725.95-461928.8 at redshift of 0.319 , which is likely the host galaxy of the GRB 210121A. 
\item With $z=0.319$, one can derive the up-limit of the initial fireball radius by Equation (2), which gives $r_0 \le 5.4 \times 10^7$ cm. Such an up-limit is full consistent with the standard fireball model.

\end{enumerate}

\acknowledgments

 B.B.Z acknowledges support by Fundamental Research Funds for the Central Universities
 (14380046), the National Key Research and Development Programs of China
 (2018YFA0404204), the National Natural Science Foundation of China (Grant
 Nos. 11833003, U2038105), the science research grants from the China Manned Space Project with NO.CMS-CSST-2021-B11, and the Program for Innovative Talents, Entrepreneur in Jiangsu. S. Xiao, X. Y. Song and S. L. Xiong acknowledge supports from the Strategic Priority Research Program on Space Science, the Chinese Academy of Sciences (Grant Nos. XDB23040400, XDA15052700). M.Zeng acknowledges funding support from the Tsinghua University Initiative Scientific Research Program. S. L. Xiong acknowledges helpful discussions with Y. F. Xu and X. L. Fan. as well as the usage of public data of Konus-Wind. B.B.Z. thanks B. Zhang for the helpful comments on the paper. This work made use of the data from the {\it Insight}-HXMT mission, a project funded by China National Space Administration (CNSA) and the Chinese Academy of Sciences (CAS). GECAM is a mission funded by the Chinese Academy of Sciences (CAS) under the Strategic Priority Research Program on Space Science. Y.Z.M. is supported by the National Postdoctoral Program for Innovative Talents (grant No. BX20200164).


\begin{thebibliography}{}
\expandafter\ifx\csname natexlab\endcsname\relax\def\natexlab#1{#1}\fi
\providecommand{\url}[1]{\href{#1}{#1}}
\providecommand{\dodoi}[1]{doi:~\href{http://doi.org/#1}{\nolinkurl{#1}}}
\providecommand{\doeprint}[1]{\href{http://ascl.net/#1}{\nolinkurl{http://ascl.net/#1}}}
\providecommand{\doarXiv}[1]{\href{https://arxiv.org/abs/#1}{\nolinkurl{https://arxiv.org/abs/#1}}}

\bibitem[{{Amati} {et~al.}(2002){Amati}, {Frontera}, {Tavani}, {in't Zand},
 {Antonelli}, {Costa}, {Feroci}, {Guidorzi}, {Heise}, {Masetti}, {Montanari},
 {Nicastro}, {Palazzi}, {Pian}, {Piro}, \& {Soffitta}}]{Amati2002}
{Amati}, L., {Frontera}, F., {Tavani}, M., {et~al.} 2002, \aap, 390, 81,
 \dodoi{10.1051/0004-6361:20020722}

\bibitem[{{Arnaud}(1996)}]{1996ASPC..101...17A}
{Arnaud}, K.~A. 1996, in Astronomical Society of the Pacific Conference Series,
 Vol. 101, Astronomical Data Analysis Software and Systems V, ed. G.~H.
 {Jacoby} \& J.~{Barnes}, 17

\bibitem[{{Atwood} {et~al.}(2009){Atwood}, {Abdo}, {Ackermann}, {Althouse},
 {Anderson}, {Axelsson}, {Baldini}, {Ballet}, {Band}, {Barbiellini},
 {Bartelt}, {Bastieri}, {Baughman}, {Bechtol}, {B{\'e}d{\'e}r{\`e}de},
 {Bellardi}, {Bellazzini}, {Berenji}, {Bignami}, {Bisello}, {Bissaldi},
 {Blandford}, {Bloom}, {Bogart}, {Bonamente}, {Bonnell}, {Borgland},
 {Bouvier}, {Bregeon}, {Brez}, {Brigida}, {Bruel}, {Burnett}, {Busetto},
 {Caliandro}, {Cameron}, {Caraveo}, {Carius}, {Carlson}, {Casandjian},
 {Cavazzuti}, {Ceccanti}, {Cecchi}, {Charles}, {Chekhtman}, {Cheung},
 {Chiang}, {Chipaux}, {Cillis}, {Ciprini}, {Claus}, {Cohen-Tanugi},
 {Condamoor}, {Conrad}, {Corbet}, {Corucci}, {Costamante}, {Cutini}, {Davis},
 {Decotigny}, {DeKlotz}, {Dermer}, {de Angelis}, {Digel}, {do Couto e Silva},
 {Drell}, {Dubois}, {Dumora}, {Edmonds}, {Fabiani}, {Farnier}, {Favuzzi},
 {Flath}, {Fleury}, {Focke}, {Funk}, {Fusco}, {Gargano}, {Gasparrini},
 {Gehrels}, {Gentit}, {Germani}, {Giebels}, {Giglietto}, {Giommi}, {Giordano},
 {Glanzman}, {Godfrey}, {Grenier}, {Grondin}, {Grove}, {Guillemot}, {Guiriec},
 {Haller}, {Harding}, {Hart}, {Hays}, {Healey}, {Hirayama}, {Hjalmarsdotter},
 {Horn}, {Hughes}, {J{\'o}hannesson}, {Johansson}, {Johnson}, {Johnson},
 {Johnson}, {Johnson}, {Kamae}, {Katagiri}, {Kataoka}, {Kavelaars}, {Kawai},
 {Kelly}, {Kerr}, {Klamra}, {Kn{\"o}dlseder}, {Kocian}, {Komin}, {Kuehn},
 {Kuss}, {Landriu}, {Latronico}, {Lee}, {Lee}, {Lemoine-Goumard}, {Lionetto},
 {Longo}, {Loparco}, {Lott}, {Lovellette}, {Lubrano}, {Madejski}, {Makeev},
 {Marangelli}, {Massai}, {Mazziotta}, {McEnery}, {Menon}, {Meurer},
 {Michelson}, {Minuti}, {Mirizzi}, {Mitthumsiri}, {Mizuno}, {Moiseev},
 {Monte}, {Monzani}, {Moretti}, {Morselli}, {Moskalenko}, {Murgia},
 {Nakamori}, {Nishino}, {Nolan}, {Norris}, {Nuss}, {Ohno}, {Ohsugi}, {Omodei},
 {Orlando}, {Ormes}, {Paccagnella}, {Paneque}, {Panetta}, {Parent}, {Pearce},
 {Pepe}, {Perazzo}, {Pesce-Rollins}, {Picozza}, {Pieri}, {Pinchera}, {Piron},
 {Porter}, {Poupard}, {Rain{\`o}}, {Rando}, {Rapposelli}, {Razzano}, {Reimer},
 {Reimer}, {Reposeur}, {Reyes}, {Ritz}, {Rochester}, {Rodriguez}, {Romani},
 {Roth}, {Russell}, {Ryde}, {Sabatini}, {Sadrozinski}, {Sanchez}, {Sander},
 {Sapozhnikov}, {Parkinson}, {Scargle}, {Schalk}, {Scolieri}, {Sgr{\`o}},
 {Share}, {Shaw}, {Shimokawabe}, {Shrader}, {Sierpowska-Bartosik}, {Siskind},
 {Smith}, {Smith}, {Spandre}, {Spinelli}, {Starck}, {Stephens}, {Strickman},
 {Strong}, {Suson}, {Tajima}, {Takahashi}, {Takahashi}, {Tanaka}, {Tenze},
 {Tether}, {Thayer}, {Thayer}, {Thompson}, {Tibaldo}, {Tibolla}, {Torres},
 {Tosti}, {Tramacere}, {Turri}, {Usher}, {Vilchez}, {Vitale}, {Wang},
 {Watters}, {Winer}, {Wood}, {Ylinen}, \& {Ziegler}}]{2009ApJ...697.1071A}
{Atwood}, W.~B., {Abdo}, A.~A., {Ackermann}, M., {et~al.} 2009, \apj, 697,
 1071, \dodoi{10.1088/0004-637X/697/2/1071}

\bibitem[{{Band}(1997)}]{Band97ApJ}
{Band}, D.~L. 1997, \apj, 486, 928, \dodoi{10.1086/304566}

\bibitem[{Bhat {et~al.}(2016)Bhat, Meegan, von Kienlin, Paciesas, Briggs,
 Burgess, Burns, Chaplin, Cleveland, Collazzi, Connaughton, Diekmann,
 Fitzpatrick, Gibby, Giles, Goldstein, Greiner, Jenke, Kippen, Kouveliotou,
 Mailyan, McBreen, Pelassa, Preece, Roberts, Sparke, Stanbro, Veres,
 Wilson-Hodge, Xiong, Younes, Yu, \& Zhang}]{Bhat_2016}
Bhat, P.~N., Meegan, C.~A., von Kienlin, A., {et~al.} 2016, The Astrophysical
 Journal Supplement Series, 223, 28, \dodoi{10.3847/0067-0049/223/2/28}

\bibitem[{{Bilicki} {et~al.}(2016){Bilicki}, {Peacock}, {Jarrett}, {Cluver},
 {Maddox}, {Brown}, {Taylor}, {Hambly}, {Solarz}, {Holwerda}, {Baldry},
 {Loveday}, {Moffett}, {Hopkins}, {Driver}, {Alpaslan}, \&
 {Bland-Hawthorn}}]{2016ApJS..225....5B}
{Bilicki}, M., {Peacock}, J.~A., {Jarrett}, T.~H., {et~al.} 2016, \apjs, 225,
 5, \dodoi{10.3847/0067-0049/225/1/5}

\bibitem[{{Chen} {et~al.}(2005){Chen}, {Lou}, {Wu}, {Qu}, {Jia}, \&
 {Yang}}]{Chen05ApJ}
{Chen}, L., {Lou}, Y.-Q., {Wu}, M., {et~al.} 2005, \apj, 619, 983,
 \dodoi{10.1086/426774}

\bibitem[{{Chen} {et~al.}(2020){Chen}, {Li}, {Huang}, {Duan}, {Zhang}, {Guo},
 {Sun}, {Wang}, {Song}, {Li}, {Li}, {Ou}, {Zhao}, {Peng}, {Shi}, {Li}, {Li},
 {Xiao}, {Song}, {Wang}, {Ma}, {Zhang}, {Xiong}, {Cai}, {Zhang}, {Chen},
 {Qiao}, {Yao}, {Zheng}, \& {Zhao}}]{2020SSPMA..50l9512C}
{Chen}, W., {Li}, B., {Huang}, Y., {et~al.} 2020, Scientia Sinica Physica,
 Mechanica \& Astronomica, 50, 129512, \dodoi{10.1360/SSPMA-2020-0389}

\bibitem[{{Du} {et~al.}(2021){Du}, {Lan}, {Wei}, {Zhou}, {Gao}, {Jiang},
 {Zhang}, {Liu}, {Wu}, {Liang}, \& {Zhu}}]{Du21}
{Du}, S.-S., {Lan}, L., {Wei}, J.-J., {et~al.} 2021, \apj, 906, 8,
 \dodoi{10.3847/1538-4357/abc624}

\bibitem[{Gao \& Zhang(2015)}]{Gao_2015}
Gao, H., \& Zhang, B. 2015, The Astrophysical Journal, 801, 103,
 \dodoi{10.1088/0004-637x/801/2/103}

\bibitem[{{Geng} {et~al.}(2018){Geng}, {Huang}, {Wu}, {Zhang}, \&
 {Zong}}]{2018ApJS..234....3G}
{Geng}, J.-J., {Huang}, Y.-F., {Wu}, X.-F., {Zhang}, B., \& {Zong}, H.-S. 2018,
 \apjs, 234, 3, \dodoi{10.3847/1538-4365/aa9e84}

\bibitem[{Gruber {et~al.}(2014)Gruber, Goldstein, von Ahlefeld, Bhat, Bissaldi,
 Briggs, Byrne, Cleveland, Connaughton, Diehl, Fishman, Fitzpatrick, Foley,
 Gibby, Giles, Greiner, Guiriec, van~der Horst, von Kienlin, Kouveliotou,
 Layden, Lin, Meegan, McGlynn, Paciesas, Pelassa, Preece, Rau, Wilson-Hodge,
 Xiong, Younes, \& Yu}]{Gruber_2014}
Gruber, D., Goldstein, A., von Ahlefeld, V.~W., {et~al.} 2014, The
 Astrophysical Journal Supplement Series, 211, 12,
 \dodoi{10.1088/0067-0049/211/1/12}

\bibitem[{{Guiriec} {et~al.}(2011){Guiriec}, {Connaughton}, {Briggs},
 {Burgess}, {Ryde}, {Daigne}, {M{\'e}sz{\'a}ros}, {Goldstein}, {McEnery},
 {Omodei}, {Bhat}, {Bissaldi}, {Camero-Arranz}, {Chaplin}, {Diehl}, {Fishman},
 {Foley}, {Gibby}, {Giles}, {Greiner}, {Gruber}, {von Kienlin}, {Kippen},
 {Kouveliotou}, {McBreen}, {Meegan}, {Paciesas}, {Preece}, {Rau}, {Tierney},
 {van der Horst}, \& {Wilson-Hodge}}]{2011ApJ...727L..33G}
{Guiriec}, S., {Connaughton}, V., {Briggs}, M.~S., {et~al.} 2011, \apjl, 727,
 L33, \dodoi{10.1088/2041-8205/727/2/L33}

\bibitem[{{Hou} {et~al.}(2018){Hou}, {Zhang}, {Meng}, {Wu}, {Liang}, {L{\"u}},
 {Liu}, {Liang}, {Lin}, {Lu}, {Huang}, \& {Zhang}}]{2018ApJ...866...13H}
{Hou}, S.-J., {Zhang}, B.-B., {Meng}, Y.-Z., {et~al.} 2018, \apj, 866, 13,
 \dodoi{10.3847/1538-4357/aadc07}

\bibitem[{{Hurley} {et~al.}(2021){Hurley}, {Ipn}, {Mitrofanov}, {Golovin},
 {Kozyrev}, {Litvak}, {Sanin}, {Hend-Odyssey Grb Team}, {Svinkin},
 {Frederiks}, {Ridnaia}, {Cline}, {Konus-Wind Team}, {Zhang}, {Rau},
 {Savchenko}, {Bozzo}, {Ferrigno}, {INTEGRAL SPI-ACS Grb Team}, {Boynton},
 {Fellows}, {Harshman}, {Enos}, {Starr}, \& {Grs-Odyssey Grb
 Team}}]{2021GCN.29348....1H}
{Hurley}, K., {Ipn}, {Mitrofanov}, I.~G., {et~al.} 2021, GRB Coordinates
 Network, 29348, 1

\bibitem[{{Ioka} \& {Nakamura}(2001)}]{Ioka01ApJL}
{Ioka}, K., \& {Nakamura}, T. 2001, \apjl, 554, L163, \dodoi{10.1086/321717}

\bibitem[{{Li}(2007)}]{Li:2007NuPhS}
{Li}, T.-P. 2007, Nuclear Physics B Proceedings Supplements, 166, 131,
 \dodoi{10.1016/j.nuclphysbps.2006.12.070}

\bibitem[{{Lu} {et~al.}(2006){Lu}, {Qin}, {Zhang}, \& {Yi}}]{Lu06MNRAS}
{Lu}, R.~J., {Qin}, Y.~P., {Zhang}, Z.~B., \& {Yi}, T.~F. 2006, \mnras, 367,
 275, \dodoi{10.1111/j.1365-2966.2005.09951.x}

\bibitem[{{Lu} {et~al.}(2012){Lu}, {Wei}, {Liang}, {Zhang}, {L{\"u}}, {L{\"u}},
 {Lei}, \& {Zhang}}]{2012ApJ...756..112L}
{Lu}, R.-J., {Wei}, J.-J., {Liang}, E.-W., {et~al.} 2012, \apj, 756, 112,
 \dodoi{10.1088/0004-637X/756/2/112}

\bibitem[{{Lundman} {et~al.}(2013){Lundman}, {Pe'er}, \&
 {Ryde}}]{2013MNRAS.428.2430L}
{Lundman}, C., {Pe'er}, A., \& {Ryde}, F. 2013, \mnras, 428, 2430,
 \dodoi{10.1093/mnras/sts219}

\bibitem[{{Luo} {et~al.}(2020){Luo}, {Liao}, {Li}, {Li}, {Zhang}, {Liu}, {Li},
 {Zhu}, {Li}, {Huang}, {Ge}, {Xu}, {Li}, {Cai}, {Xiao}, {Yi}, {Zhang},
 {Xiong}, {Zhang}, \& {Zhang}}]{2020JHEAp..27....1L}
{Luo}, Q., {Liao}, J.-Y., {Li}, X.-F., {et~al.} 2020, Journal of High Energy
 Astrophysics, 27, 1, \dodoi{10.1016/j.jheap.2020.04.004}

\bibitem[{{Meegan} {et~al.}(2009){Meegan}, {Lichti}, {Bhat}, {Bissaldi},
 {Briggs}, {Connaughton}, {Diehl}, {Fishman}, {Greiner}, {Hoover}, {van der
 Horst}, {von Kienlin}, {Kippen}, {Kouveliotou}, {McBreen}, {Paciesas},
 {Preece}, {Steinle}, {Wallace}, {Wilson}, \&
 {Wilson-Hodge}}]{2009ApJ...702..791M}
{Meegan}, C., {Lichti}, G., {Bhat}, P.~N., {et~al.} 2009, \apj, 702, 791,
 \dodoi{10.1088/0004-637X/702/1/791}

\bibitem[{{Meng} {et~al.}(2021){Meng}, {Geng}, \& {Wu}}]{2021arXiv210704532M}
{Meng}, Y.-Z., {Geng}, J.-J., \& {Wu}, X.-F. 2021, arXiv e-prints,
 arXiv:2107.04532.
\newblock \doarXiv{2107.04532}

\bibitem[{{Meng} {et~al.}(2019){Meng}, {Liu}, {Wei}, {Wu}, \&
 {Zhang}}]{2019ApJ...882...26M}
{Meng}, Y.-Z., {Liu}, L.-D., {Wei}, J.-J., {Wu}, X.-F., \& {Zhang}, B.-B. 2019,
 \apj, 882, 26, \dodoi{10.3847/1538-4357/ab30c7}

\bibitem[{{Meng} {et~al.}(2018){Meng}, {Geng}, {Zhang}, {Wei}, {Xiao}, {Liu},
 {Gao}, {Wu}, {Liang}, {Huang}, {Dai}, \& {Zhang}}]{2018ApJ...860...72M}
{Meng}, Y.-Z., {Geng}, J.-J., {Zhang}, B.-B., {et~al.} 2018, \apj, 860, 72,
 \dodoi{10.3847/1538-4357/aac2d9}

\bibitem[{{M{\'e}sz{\'a}ros} {et~al.}(2002){M{\'e}sz{\'a}ros}, {Ramirez-Ruiz},
 {Rees}, \& {Zhang}}]{2002ApJ...578..812M}
{M{\'e}sz{\'a}ros}, P., {Ramirez-Ruiz}, E., {Rees}, M.~J., \& {Zhang}, B. 2002,
 \apj, 578, 812, \dodoi{10.1086/342611}

\bibitem[{{M{\'e}sz{\'a}ros} \& {Rees}(2000)}]{2000ApJ...530..292M}
{M{\'e}sz{\'a}ros}, P., \& {Rees}, M.~J. 2000, \apj, 530, 292,
 \dodoi{10.1086/308371}

\bibitem[{{Norris} {et~al.}(2000{\natexlab{a}}){Norris}, {Marani}, \&
 {Bonnell}}]{Norris00ApJ}
{Norris}, J.~P., {Marani}, G.~F., \& {Bonnell}, J.~T. 2000{\natexlab{a}}, \apj,
 534, 248, \dodoi{10.1086/308725}

\bibitem[{{Norris} {et~al.}(2000{\natexlab{b}}){Norris}, {Marani}, \&
 {Bonnell}}]{2000ApJ...534..248N}
---. 2000{\natexlab{b}}, \apj, 534, 248, \dodoi{10.1086/308725}

\bibitem[{{Norris} {et~al.}(1996){Norris}, {Nemiroff}, {Bonnell}, {Scargle},
 {Kouveliotou}, {Paciesas}, {Meegan}, \& {Fishman}}]{norris96}
{Norris}, J.~P., {Nemiroff}, R.~J., {Bonnell}, J.~T., {et~al.} 1996, \apj, 459,
 393, \dodoi{10.1086/176902}

\bibitem[{{Norris} {et~al.}(1986){Norris}, {Share}, {Messina}, {Dennis},
 {Desai}, {Cline}, {Matz}, \& {Chupp}}]{Norris86ApJ}
{Norris}, J.~P., {Share}, G.~H., {Messina}, D.~C., {et~al.} 1986, \apj, 301,
 213, \dodoi{10.1086/163889}

\bibitem[{{Pe'er}(2008)}]{2008ApJ...682..463P}
{Pe'er}, A. 2008, \apj, 682, 463, \dodoi{10.1086/588136}

\bibitem[{{Pe'er} {et~al.}(2015){Pe'er}, {Barlow}, {O'Mahony}, {Margutti},
 {Ryde}, {Larsson}, {Lazzati}, \& {Livio}}]{2015ApJ...813..127P}
{Pe'er}, A., {Barlow}, H., {O'Mahony}, S., {et~al.} 2015, \apj, 813, 127,
 \dodoi{10.1088/0004-637X/813/2/127}

\bibitem[{{Peng} {et~al.}(2021{\natexlab{a}}){Peng}, {Xiao}, {Cai}, {Zhao},
 {Wang}, {Xiong}, {An}, {Chen}, {Chen}, {Chen}, {Gao}, {Gong}, {Guo}, {He},
 {Huang}, {Li}, {Li}, {Li}, {Li}, {Li}, {Li}, {Li}, {Li}, {Liang}, {Liao},
 {Liu}, {Liu}, {Liu}, {Lu}, {Luo}, {Ma}, {Ou}, {Qiao}, {Shi}, {Shi}, {Song},
 {Song}, {Sun}, {Sun}, {Tuo}, {Wang}, {Wang}, {Wen}, {Xu}, {Xu}, {Xue},
 {Yang}, {Yao}, {Yi}, {Zhang}, {Zhang}, {Zhang}, {Zhang}, {Zhang}, {Zhang},
 {Zhang}, {Zhang}, {Zhang}, {Zhang}, {Zhao}, {Zhao}, {Zheng}, {Zheng}, {Zhou},
 \& {Gecam Team}}]{GCN29347}
{Peng}, W.~X., {Xiao}, S., {Cai}, C., {et~al.} 2021{\natexlab{a}}, GRB
 Coordinates Network, 29347, 1

\bibitem[{{Peng} {et~al.}(2021{\natexlab{b}}){Peng}, {Xiao}, {Cai}, {Zhao},
 {Wang}, {Xiong}, {An}, {Chen}, {Chen}, {Chen}, {Gao}, {Gong}, {Guo}, {He},
 {Huang}, {Li}, {Li}, {Li}, {Li}, {Li}, {Li}, {Li}, {Li}, {Liang}, {Liao},
 {Liu}, {Liu}, {Liu}, {Lu}, {Luo}, {Ma}, {Ou}, {Qiao}, {Shi}, {Shi}, {Song},
 {Song}, {Sun}, {Sun}, {Tuo}, {Wang}, {Wang}, {Wen}, {Xu}, {Xu}, {Xue},
 {Yang}, {Yao}, {Yi}, {Zhang}, {Zhang}, {Zhang}, {Zhang}, {Zhang}, {Zhang},
 {Zhang}, {Zhang}, {Zhang}, {Zhang}, {Zhao}, {Zhao}, {Zheng}, {Zheng}, {Zhou},
 \& {Gecam Team}}]{2021GCN.29347....1P}
---. 2021{\natexlab{b}}, GRB Coordinates Network, 29347, 1

\bibitem[{{Poolakkil} {et~al.}(2021){Poolakkil}, {Preece}, {Fletcher},
 {Goldstein}, {Bhat}, {Bissaldi}, {Briggs}, {Burns}, {Cleveland}, {Giles},
 {Hui}, {Kocevski}, {Lesage}, {Mailyan}, {Malacaria}, {Paciesas}, {Roberts},
 {Veres}, {von Kienlin}, \& {Wilson-Hodge}}]{2021arXiv210313528P}
{Poolakkil}, S., {Preece}, R., {Fletcher}, C., {et~al.} 2021, arXiv e-prints,
 arXiv:2103.13528.
\newblock \doarXiv{2103.13528}

\bibitem[{{Preece} {et~al.}(1998){Preece}, {Briggs}, {Mallozzi}, {Pendleton},
 {Paciesas}, \& {Band}}]{1998ApJ...506L..23P}
{Preece}, R.~D., {Briggs}, M.~S., {Mallozzi}, R.~S., {et~al.} 1998, \apjl, 506,
 L23, \dodoi{10.1086/311644}

\bibitem[{{Racusin} {et~al.}(2011){Racusin}, {Oates}, {Schady}, {Burrows}, {de
 Pasquale}, {Donato}, {Gehrels}, {Koch}, {McEnery}, {Piran}, {Roming},
 {Sakamoto}, {Swenson}, {Troja}, {Vasileiou}, {Virgili}, {Wanderman}, \&
 {Zhang}}]{2011ApJ...738..138R}
{Racusin}, J.~L., {Oates}, S.~R., {Schady}, P., {et~al.} 2011, \apj, 738, 138,
 \dodoi{10.1088/0004-637X/738/2/138}

\bibitem[{{Roberts} {et~al.}(2021){Roberts}, {Veres}, {Baring}, {Briggs},
 {Kouveliotou}, {Bissaldi}, {Younes}, {Chastain}, {DeLaunay}, {Huppenkothen},
 {Tohuvavohu}, {Bhat}, {G{\"o}{\v{g}}{\"u}{\c{s}}}, {van der Horst}, {Kennea},
 {Kocevski}, {Linford}, {Guiriec}, {Hamburg}, {Wilson-Hodge}, \&
 {Burns}}]{Roberts2021Natur}
{Roberts}, O.~J., {Veres}, P., {Baring}, M.~G., {et~al.} 2021, \nat, 589, 207,
 \dodoi{10.1038/s41586-020-03077-8}

\bibitem[{{Scargle} {et~al.}(2013){Scargle}, {Norris}, {Jackson}, \&
 {Chiang}}]{Scargle2013ApJ}
{Scargle}, J.~D., {Norris}, J.~P., {Jackson}, B., \& {Chiang}, J. 2013, \apj,
 764, 167, \dodoi{10.1088/0004-637X/764/2/167}

\bibitem[{{Schwarz}(1978)}]{1978AnSta...6..461S}
{Schwarz}, G. 1978, Annals of Statistics, 6, 461

\bibitem[{{Shen} {et~al.}(2005){Shen}, {Song}, \& {Li}}]{Shen05MNRAS}
{Shen}, R.-F., {Song}, L.-M., \& {Li}, Z. 2005, \mnras, 362, 59,
 \dodoi{10.1111/j.1365-2966.2005.09163.x}

\bibitem[{{Shenoy} {et~al.}(2013){Shenoy}, {Sonbas}, {Dermer}, {Maximon},
 {Dhuga}, {Bhat}, {Hakkila}, {Parke}, {Maclachlan}, {Eskandarian}, \&
 {Ukwatta}}]{Shenoy13ApJ}
{Shenoy}, A., {Sonbas}, E., {Dermer}, C., {et~al.} 2013, \apj, 778, 3,
 \dodoi{10.1088/0004-637X/778/1/3}

\bibitem[{{Svinkin} {et~al.}(2021){Svinkin}, {Frederiks}, {Hurley}, {Aptekar},
 {Golenetskii}, {Lysenko}, {Ridnaia}, {Tsvetkova}, {Ulanov}, {Cline},
 {Mitrofanov}, {Golovin}, {Kozyrev}, {Litvak}, {Sanin}, {Goldstein}, {Briggs},
 {Wilson-Hodge}, {von Kienlin}, {Zhang}, {Rau}, {Savchenko}, {Bozzo},
 {Ferrigno}, {Ubertini}, {Bazzano}, {Rodi}, {Barthelmy}, {Cummings}, {Krimm},
 {Palmer}, {Boynton}, {Fellows}, {Harshman}, {Enos}, \&
 {Starr}}]{Svinkin2021Natur}
{Svinkin}, D., {Frederiks}, D., {Hurley}, K., {et~al.} 2021, \nat, 589, 211,
 \dodoi{10.1038/s41586-020-03076-9}

\bibitem[{{Uhm} \& {Zhang}(2014)}]{2014NatPh..10..351U}
{Uhm}, Z.~L., \& {Zhang}, B. 2014, Nature Physics, 10, 351,
 \dodoi{10.1038/nphys2932}

\bibitem[{{Uhm} \& {Zhang}(2016)}]{Uhm&Zhang16}
---. 2016, \apj, 825, 97, \dodoi{10.3847/0004-637X/825/2/97}

\bibitem[{von Kienlin {et~al.}(2014)von Kienlin, Meegan, Paciesas, Bhat,
 Bissaldi, Briggs, Burgess, Byrne, Chaplin, Cleveland, Connaughton, Collazzi,
 Fitzpatrick, Foley, Gibby, Giles, Goldstein, Greiner, Gruber, Guiriec,
 van~der Horst, Kouveliotou, Layden, McBreen, McGlynn, Pelassa, Preece, Rau,
 Tierney, Wilson-Hodge, Xiong, Younes, \& Yu}]{von_Kienlin_2014}
von Kienlin, A., Meegan, C.~A., Paciesas, W.~S., {et~al.} 2014, The
 Astrophysical Journal Supplement Series, 211, 13,
 \dodoi{10.1088/0067-0049/211/1/13}

\bibitem[{{von Kienlin} {et~al.}(2020){von Kienlin}, {Meegan}, {Paciesas},
 {Bhat}, {Bissaldi}, {Briggs}, {Burns}, {Cleveland}, {Gibby}, {Giles},
 {Goldstein}, {Hamburg}, {Hui}, {Kocevski}, {Mailyan}, {Malacaria},
 {Poolakkil}, {Preece}, {Roberts}, {Veres}, \&
 {Wilson-Hodge}}]{von_Kienlin_2020}
{von Kienlin}, A., {Meegan}, C.~A., {Paciesas}, W.~S., {et~al.} 2020, \apj,
 893, 46, \dodoi{10.3847/1538-4357/ab7a18}

\bibitem[{{Wei} {et~al.}(2017){Wei}, {Zhang}, {Shao}, {Wu}, \&
 {M{\'e}sz{\'a}ros}}]{Wei17}
{Wei}, J.-J., {Zhang}, B.-B., {Shao}, L., {Wu}, X.-F., \& {M{\'e}sz{\'a}ros},
 P. 2017, \apjl, 834, L13, \dodoi{10.3847/2041-8213/834/2/L13}

\bibitem[{{Wen} {et~al.}(2019){Wen}, {Long}, {Zheng}, {An}, {Cai}, {Cang},
 {Che}, {Chen}, {Chen}, {Chen}, {Chen}, {Cheng}, {Deng}, {Deng}, {Ding}, {Du},
 {Duan}, {Gan}, {Gao}, {Gao}, {Han}, {Han}, {He}, {He}, {Hou}, {Hu}, {Hu},
 {Huang}, {Huang}, {Huang}, {Jia}, {Jiang}, {Jin}, {Li}, {Li}, {Li}, {Liang},
 {Liang}, {Lin}, {Liu}, {Liu}, {Liu}, {Liu}, {Liu}, {Liu}, {Lu}, {Lu}, {Lu},
 {Luo}, {Ma}, {Ma}, {Mao}, {Mo}, {Nie}, {Qu}, {Shan}, {Shi}, {Song}, {Sun},
 {Tan}, {Tang}, {Tao}, {Wang}, {Wang}, {Wang}, {Wu}, {Wu}, {Xia}, {Xiao},
 {Xie}, {Xu}, {Xu}, {Xu}, {Yan}, {Yan}, {Yang}, {Yang}, {Yang}, {Yang},
 {Yang}, {Yao}, {Yu}, {Yu}, {Zhang}, {Zhang}, {Zhang}, {Zhang}, {Zhang},
 {Zhang}, {Zhang}, {Zhao}, {Zhao}, {Zheng}, {Zhou}, {Zhu}, {Zou}, {An}, {Cai},
 {Chen}, {Dai}, {Fan}, {Feng}, {Feng}, {Gao}, {Huang}, {Kang}, {Li}, {Li},
 {Liang}, {Lin}, {Lin}, {Liu}, {Liu}, {Liu}, {Liu}, {Lu}, {Mao}, {Shen},
 {Shu}, {Su}, {Sun}, {Tam}, {Tang}, {Tian}, {Wang}, {Wang}, {Wang}, {Wang},
 {Wu}, {Wu}, {Xiong}, {Xu}, {Yu}, {Yu}, {Yu}, {Zeng}, {Zeng}, {Zhang},
 {Zhang}, {Zhao}, {Zhou}, \& {Zhu}}]{2019ExA....48...77W}
{Wen}, J., {Long}, X., {Zheng}, X., {et~al.} 2019, Experimental Astronomy, 48,
 77, \dodoi{10.1007/s10686-019-09636-w}

\bibitem[{{Xue} {et~al.}(2021{\natexlab{a}}){Xue}, {Cai}, {Liu}, {Luo}, {Xiao},
 {Yi}, {Zhang}, {Zheng}, {Huang}, {Li}, {Li}, {Li}, {Liao}, {Song}, {Xiong},
 {Liu}, {Li}, {Li}, {Chang}, {Zhang}, {Zhang}, {Lu}, {Zou}, {Jin}, {Zhang},
 {Li}, {Lu}, {Song}, {Wu}, {Xu}, {Zhang}, \& {Insight-HXMT Team}}]{GCN.29346}
{Xue}, W.~C., {Cai}, C., {Liu}, J.~C., {et~al.} 2021{\natexlab{a}}, GRB
 Coordinates Network, 29346, 1

\bibitem[{{Xue} {et~al.}(2021{\natexlab{b}}){Xue}, {Cai}, {Liu}, {Luo}, {Xiao},
 {Yi}, {Zhang}, {Zheng}, {Huang}, {Li}, {Li}, {Li}, {Liao}, {Song}, {Xiong},
 {Liu}, {Li}, {Li}, {Chang}, {Zhang}, {Zhang}, {Lu}, {Zou}, {Jin}, {Zhang},
 {Li}, {Lu}, {Song}, {Wu}, {Xu}, {Zhang}, \& {Insight-HXMT
 Team}}]{2021GCN.29346....1X}
---. 2021{\natexlab{b}}, GRB Coordinates Network, 29346, 1

\bibitem[{{Yang} {et~al.}(2020{\natexlab{a}}){Yang}, {Chand}, {Zhang}, {Yang},
 {Zou}, {Yang}, {Zhao}, {Shao}, {Xiong}, {Luo}, {Li}, {Xiao}, {Li}, {Liu},
 {Joshi}, {Sharma}, {Chakraborty}, {Li}, \& {Zhang}}]{JunYang2020ApJ}
{Yang}, J., {Chand}, V., {Zhang}, B.-B., {et~al.} 2020{\natexlab{a}}, \apj,
 899, 106, \dodoi{10.3847/1538-4357/aba745}

\bibitem[{{Yang} {et~al.}(2020{\natexlab{b}}){Yang}, {Zhong}, {Zhang}, {Wu},
 {Zhang}, {Yang}, {Cao}, {Gao}, {Zou}, {Wang}, {L{\"u}}, {Cang}, \&
 {Dai}}]{Y.S.Yang2020ApJ}
{Yang}, Y.-S., {Zhong}, S.-Q., {Zhang}, B.-B., {et~al.} 2020{\natexlab{b}},
 \apj, 899, 60, \dodoi{10.3847/1538-4357/ab9ff5}

\bibitem[{{Yi} {et~al.}(2006){Yi}, {Liang}, {Qin}, \& {Lu}}]{yi06}
{Yi}, T., {Liang}, E., {Qin}, Y., \& {Lu}, R. 2006, \mnras, 367, 1751,
 \dodoi{10.1111/j.1365-2966.2006.10083.x}

\bibitem[{{Yu} {et~al.}(2016){Yu}, {Preece}, {Greiner}, {Narayana Bhat},
 {Bissaldi}, {Briggs}, {Cleveland}, {Connaughton}, {Goldstein}, {von Kienlin},
 {Kouveliotou}, {Mailyan}, {Meegan}, {Paciesas}, {Rau}, {Roberts}, {Veres},
 {Wilson-Hodge}, {Zhang}, \& {van Eerten}}]{2016A&A...588A.135Y}
{Yu}, H.-F., {Preece}, R.~D., {Greiner}, J., {et~al.} 2016, \aap, 588, A135,
 \dodoi{10.1051/0004-6361/201527509}

\bibitem[{{Zhang} {et~al.}(2012{\natexlab{a}}){Zhang}, {Lu}, {Liang}, \&
 {Wu}}]{2012ApJ...758L..34Z}
{Zhang}, B., {Lu}, R.-J., {Liang}, E.-W., \& {Wu}, X.-F. 2012{\natexlab{a}},
 \apjl, 758, L34, \dodoi{10.1088/2041-8205/758/2/L34}

\bibitem[{Zhang \& Pe{\textquotesingle}er(2009)}]{Zhang_2009}
Zhang, B., \& Pe{\textquotesingle}er, A. 2009, The Astrophysical Journal, 700,
 L65, \dodoi{10.1088/0004-637x/700/2/l65}

\bibitem[{{Zhang} \& {Yan}(2011)}]{2011ApJ...726...90Z}
{Zhang}, B., \& {Yan}, H. 2011, \apj, 726, 90,
 \dodoi{10.1088/0004-637X/726/2/90}

\bibitem[{{Zhang} {et~al.}(2016){Zhang}, {Uhm}, {Connaughton}, {Briggs}, \&
 {Zhang}}]{2016ApJ...816...72Z}
{Zhang}, B.-B., {Uhm}, Z.~L., {Connaughton}, V., {Briggs}, M.~S., \& {Zhang},
 B. 2016, \apj, 816, 72, \dodoi{10.3847/0004-637X/816/2/72}

\bibitem[{{Zhang} {et~al.}(2012{\natexlab{b}}){Zhang}, {Burrows}, {Zhang},
 {M{\'e}sz{\'a}ros}, {Wang}, {Stratta}, {D'Elia}, {Frederiks}, {Golenetskii},
 {Cummings}, {Norris}, {Falcone}, {Barthelmy}, \&
 {Gehrels}}]{2012ApJ...748..132Z}
{Zhang}, B.-B., {Burrows}, D.~N., {Zhang}, B., {et~al.} 2012{\natexlab{b}},
 \apj, 748, 132, \dodoi{10.1088/0004-637X/748/2/132}

\bibitem[{{Zhang} {et~al.}(2018{\natexlab{a}}){Zhang}, {Zhang},
 {Castro-Tirado}, {Dai}, {Tam}, {Wang}, {Hu}, {Karpov}, {Pozanenko}, {Zhang},
 {Mazaeva}, {Minaev}, {Volnova}, {Oates}, {Gao}, {Wu}, {Shao}, {Tang},
 {Beskin}, {Biryukov}, {Bondar}, {Ivanov}, {Katkova}, {Orekhova}, {Perkov},
 {Sasyuk}, {Mankiewicz}, {{\.Z}arnecki}, {Cwiek}, {Opiela}, {Zadro{\.Z}ny},
 {Aptekar}, {Frederiks}, {Svinkin}, {Kusakin}, {Inasaridze}, {Burhonov},
 {Rumyantsev}, {Klunko}, {Moskvitin}, {Fatkhullin}, {Sokolov}, {Valeev},
 {Jeong}, {Park}, {Caballero-Garc{\'\i}a}, {Cunniffe}, {Tello}, {Ferrero},
 {Pandey}, {Jel{\'\i}nek}, {Peng}, {S{\'a}nchez-Ram{\'\i}rez}, \&
 {Castell{\'o}n}}]{2018NatAs...2...69Z}
{Zhang}, B.~B., {Zhang}, B., {Castro-Tirado}, A.~J., {et~al.}
 2018{\natexlab{a}}, Nature Astronomy, 2, 69,
 \dodoi{10.1038/s41550-017-0309-8}

\bibitem[{{Zhang} {et~al.}(2020{\natexlab{a}}){Zhang}, {Liu}, {Zhong}, \&
 {Wang}}]{H.M.Zhang2020ApJ}
{Zhang}, H.-M., {Liu}, R.-Y., {Zhong}, S.-Q., \& {Wang}, X.-Y.
 2020{\natexlab{a}}, \apjl, 903, L32, \dodoi{10.3847/2041-8213/abc2c9}

\bibitem[{{Zhang} {et~al.}(2018{\natexlab{b}}){Zhang}, {Zhang}, {Lu}, {Li},
 {Song}, {Xu}, {Wang}, {Qu}, {Liu}, {Chen}, {Cao}, {Zhang}, {Xiong}, {Ge},
 {Chen}, {Liao}, {Nie}, {Zhao}, {Jia}, {Li}, {Guan}, {Li}, {Zhang}, {Jin},
 {Wang}, {Zheng}, {Ma}, {Tao}, \& {Huang}}]{2018SPIE10699E..1UZ}
{Zhang}, S., {Zhang}, S.~N., {Lu}, F.~J., {et~al.} 2018{\natexlab{b}}, in
 Society of Photo-Optical Instrumentation Engineers (SPIE) Conference Series,
 Vol. 10699, Space Telescopes and Instrumentation 2018: Ultraviolet to Gamma
 Ray, ed. J.-W.~A. {den Herder}, S.~{Nikzad}, \& K.~{Nakazawa}, 106991U,
 \dodoi{10.1117/12.2311835}

\bibitem[{{Zhang} {et~al.}(2020{\natexlab{b}}){Zhang}, {Li}, {Lu}, {Song},
 {Xu}, {Liu}, {Chen}, {Cao}, {Bu}, {Chang}, {Chen}, {Chen}, {Chen}, {Chen},
 {Chen}, {Cui}, {Cui}, {Deng}, {Dong}, {Du}, {Fu}, {Gao}, {Gao}, {Gao}, {Ge},
 {Gu}, {Guan}, {Gungor}, {Guo}, {Han}, {Hu}, {Huang}, {Huo}, {Jia}, {Jiang},
 {Jiang}, {Jin}, {Jin}, {Li}, {Li}, {Li}, {Li}, {Li}, {Li}, {Li}, {Li}, {Li},
 {Li}, {Li}, {Liang}, {Liao}, {Liu}, {Liu}, {Liu}, {Liu}, {Liu}, {Liu}, {Lu},
 {Lu}, {Luo}, {Ma}, {Meng}, {Nang}, {Nie}, {Ou}, {Qu}, {Sai}, {Shang}, {Shen},
 {Sun}, {Tan}, {Tao}, {Tuo}, {Wang}, {Wang}, {Wang}, {Wang}, {Wang}, {Wang},
 {Wang}, {Wen}, {Wu}, {Wu}, {Wu}, {Xiao}, {Xiong}, {Yan}, {Yang}, {Yang},
 {Yang}, {Yi}, {Yuan}, {Zhang}, {Zhang}, {Zhang}, {Zhang}, {Zhang}, {Zhang},
 {Zhang}, {Zhang}, {Zhang}, {Zhang}, {Zhang}, {Zhang}, {Zhang}, {Zhang},
 {Zhang}, {Zhang}, {Zhang}, {Zhang}, {Zhang}, {Zhang}, {Zhao}, {Zhao},
 {Zheng}, {Zhou}, {Zhu}, {Zhu}, {Zhuang}, \& {Insight-HXMT
 team}}]{Zhang:2020SCPMA}
{Zhang}, S.-N., {Li}, T., {Lu}, F., {et~al.} 2020{\natexlab{b}}, Science China
 Physics, Mechanics, and Astronomy, 63, 249502,
 \dodoi{10.1007/s11433-019-1432-6}




\end{thebibliography}

\end{document}